\documentclass[12pt]{article}
\pdfoutput=1
\usepackage{graphicx,amssymb,amsmath,empheq}
\usepackage{authblk}
\usepackage{amsthm,bm}
\usepackage{empheq}
\usepackage{subfig} 

\usepackage{hyperref}
\hypersetup{colorlinks = true, linkcolor=black, citecolor=black, urlcolor=blue}
\usepackage{url}

\theoremstyle{plain}

\usepackage{tikz}
\usepackage{tikz-cd}
\usetikzlibrary{arrows}
\usetikzlibrary{intersections}
\usetikzlibrary{shapes.geometric}
\usetikzlibrary{decorations.pathmorphing, patterns,shapes}
\usetikzlibrary{decorations.markings}

\tikzset{
  mid arrow/.style={postaction={decorate,decoration={
        markings,
        mark=at position .575 with {\arrow[#1]{stealth}}
      }}},
  near arrow/.style={postaction={decorate,decoration={
        markings,
        mark=at position .275 with {\arrow[#1]{stealth}}
      }}},
   far arrow/.style={postaction={decorate,decoration={
        markings,
        mark=at position .800 with {\arrow[#1]{stealth}}
      }}},
}

\tikzset{
  baseline = -0.5ex,
  wavy/.style = {
    thick,
    decorate,
    decoration={snake,amplitude=2pt,segment length=5pt}},
  sdot/.style = {
    circle,
    draw=none,
    fill=black,
    minimum size=2.5pt,
    inner sep=0pt},
  bdot/.style = {
    circle,
    draw=none,
    fill=black,
    minimum size=4pt,
    inner sep=0pt},
  svertex/.style = {
    circle,
    draw=black,
    thick,
    fill=lightgray,
    minimum size=8pt,
    inner sep=1pt},
  mvertex/.style = {
    circle,
    draw=black,
    thick,
    fill=lightgray,
    minimum size=12pt,
    inner sep=1pt},
  bvertex/.style = {
    circle,
    draw=black,
    thick,
    fill=lightgray,
    minimum size=16pt}}

\usepackage[nosort]{cite}

\pgfdeclarepatternformonly{south west lines}{\pgfqpoint{-0pt}{-0pt}}{\pgfqpoint{3pt}{3pt}}{\pgfqpoint{3pt}{3pt}}{
	\pgfsetlinewidth{0.4pt}
	\pgfpathmoveto{\pgfqpoint{0pt}{0pt}}
	\pgfpathlineto{\pgfqpoint{3pt}{3pt}}
	\pgfpathmoveto{\pgfqpoint{2.8pt}{-.2pt}}
	\pgfpathlineto{\pgfqpoint{3.2pt}{.2pt}}
	\pgfpathmoveto{\pgfqpoint{-.2pt}{2.8pt}}
	\pgfpathlineto{\pgfqpoint{.2pt}{3.2pt}}
	\pgfusepath{stroke}}

\tikzset{
	mid arrow/.style={postaction={decorate,decoration={
				markings,
				mark=at position .575 with {\arrow{stealth}}
	}}},
	near arrow/.style={postaction={decorate,decoration={
				markings,
				mark=at position .275 with {\arrow{stealth}}
	}}},
	far arrow/.style={postaction={decorate,decoration={
				markings,
				mark=at position .800 with {\arrow{stealth}}
	}}},
	snake arrow/.style={fixed point arithmetic, decorate, decoration={snake,amplitude=2pt, segment length=11pt},postaction={decoration={markings,mark=at position 0.625 with {\arrow{stealth}}},decorate}},
}
\tikzset{
  baseline = -0.5ex,
  wavy/.style = {
    thick,
    decorate,
    decoration={snake,amplitude=2pt,segment length=5pt}},
  sdot/.style = {
    circle,
    draw=none,
    fill=black,
    minimum size=2.5pt,
    inner sep=0pt},
  bdot/.style = {
    circle,
    draw=none,
    fill=black,
    minimum size=4pt,
    inner sep=0pt},
  svertex/.style = {
    circle,
    draw=black,
    thick,
    fill=lightgray,
    minimum size=14pt,
    inner sep=1pt},
  bvertex/.style = {
    circle,
    draw=black,
    thick,
    fill=lightgray,
    minimum size=24pt},
  bvertexsmall/.style = {
    circle,
    draw=black,
    thick,
    fill=lightgray,
    minimum size=3pt},
  bvertexnormal/.style = {
    circle,
    draw=black,
    thick,
    fill=lightgray,
    minimum size=16pt},
    bvertexnormal2/.style = {
    circle,
    draw=black,
    thick,
    fill=lightgray,
    minimum size=24pt},
  dvertex/.style = {
    circle,
    draw=black,
    thick,
    fill=gray,
    minimum size=25pt}}

\makeatletter

\newcommand*{\wideboxed}[1]{\setlength{\fboxsep}{1ex}%
  \fbox{\m@th$\displaystyle#1$}}
\makeatother

\title{
Fidelity of Wormhole Teleportation in Finite-qubit Systems
}

\author[1]{Zeyu Liu}
\author[1,2]{Pengfei Zhang}
\affil[1]{\normalsize \it Department of Physics, Fudan University, Shanghai, 200438, China}
\affil[2]{\normalsize\it Shanghai Qi Zhi Institute, AI Tower, Xuhui District, Shanghai 200232, China}

\date{\today}

\begin{document}
  \maketitle

\begin{abstract}
The rapid development of quantum science and technology is leading us into an era where quantum many-body systems can be comprehended through quantum simulations. Holographic duality, which states gravity and spacetime can emerge from strongly interacting systems, then offers a natural avenue for the experimental study of gravity physics without delving into experimentally infeasible high energies. A prominent example is the simulation of traversable wormholes through the wormhole teleportation protocol, attracting both theoretical and experimental attention. In this work, we develop the theoretical framework for computing the fidelity of wormhole teleportation in $N$-qubit systems with all-to-all interactions, quantified by mutual information and entanglement negativity. The main technique is the scramblon effective theory, which captures universal out-of-time-order correlations in generic chaotic systems. We clarify that strong couplings between the two systems are essential for simulating the probe limit of semi-classical traversable wormholes using strongly interacting systems with near-maximal chaos. However, the teleportation signal diminishes rapidly when reducing the system size $N$, requiring a large number of qubits to observe a sharp signature of emergent geometry by simulating the Sachdev-Ye-Kitaev model. This includes both the causal time-order of signals and the asymmetry of the teleportation signal for coupling with different signs. As a comparison, the teleportation signal increases when reducing $N$ in weakly interacting systems. We also analyze the fidelity of the generalized encoding scheme in fermionic string operators.
  \end{abstract}

\tableofcontents

\section{Introduction}
    Experiments play a crucial role for the understanding of quantum many-body systems. They provide valuable insights, validate theoretical models, and drive advancements in both theoretical and experimental techniques. Recent years have witnessed significant breakthroughs in quantum simulation platforms, offering a valuable avenue for experimentally investigating theoretical models without the need to search for realistic materials. As a comparison, the development of a quantum theory of gravity is largely obstructed by the large Planck mass, an energy scale that is infeasible for any controlled experiment. Fortunately, the holographic duality \cite{tHooft:1993dmi,Susskind:1994vu,Maldacena:1997re}, which states a $(D+1)+1$ dimensional gravity system can be alternatively described by a $D+1$ dimensional quantum theory. This perspective facilitates the quantum simulation of intricate gravitational physics through more manageable tabletop experiments, leading us to an exciting era of ``quantum gravity in lab'' \cite{Brown:2019hmk,Nezami:2021yaq}.

    A concrete example of quantum simulation of gravity physics involves the exploration of traversable wormholes \cite{Gao:2016bin,Maldacena:2017axo} using a specific teleportation protocol inspired by holography \cite{Brown:2019hmk,Nezami:2021yaq,Maldacena:2017axo,Gao:2018yzk,Gao:2019nyj,Schuster:2021uvg}, as briefly reviewed later. Conceptually, this is natural, as quantum teleportation utilizes the resources of quantum entanglement to transfer unknown quantum states from a sender to a receiver at a different location. Meanwhile, the ER=EPR conjecture \cite{Maldacena:2013xja} proposes that entanglement can geometrize into a wormhole, which should be traversable for the quantum signal. Intriguingly, the emergence of a traversable wormhole yields sharp signatures, such as the causal ordering of signals and asymmetry of the teleportation signal for coupling $\mu$ with different signs, which can be unambiguously tested. Furthermore, as a concrete quantum mechanical model for holography, the Sachdev-Ye-Kitaev (SYK) model has been introduced \cite{Kit.KITP.2, Maldacena:2016hyu, Maldacena:2016upp}, which is related to the earlier exploration of random spin models in \cite{Sachdev:1992fk}. Putting all these advancements together, a pioneering experiment has been conducted on the Google Sycamore processor \cite{Jafferis:2022crx}. However, it is limited to a small system size with $N=7$ Majorana fermions and a specific sparse commuting SYK model \cite{Gao:2023gta}, generated by the machine learning method, due to restrictions in the fidelity of quantum gates. On the other hand, main analytical results on the quantum side are obtained in solvable systems with large local Hilbert space dimensions, including the SYK model in the large-$N$ limit \cite{Brown:2019hmk,Nezami:2021yaq,Gao:2019nyj,Schuster:2021uvg} and the conformal field theory with large central charges \cite{Gao:2018yzk}. 

    The purpose of this work is to bridge the gap between theoretical description and experiments by developing a framework to compute the fidelity of wormhole teleportation in systems with a finite number of qubits. Our primary theoretical tool is the scramblon effective theory \cite{Gu:2021xaj,Stanford:2021bhl}, known for its efficacy in computing out-of-time-order (OTO) correlations in generic chaotic many-body systems with all-to-all interactions, such as out-of-time-order correlators (OTOCs) \cite{Gu:2021xaj,Zhang:2023vpm} and operator size distributions \cite{Zhang:2022fma,Zhang:2022knu,Liu:2023lyu}. It posits that all significant correlations between operators are mediated by collective modes known as scramblons \cite{Kitaev:2017awl} for sufficiently long times. As we will explain, out-of-time-order correlations play a dominant role in wormhole teleportation. In Ref. \cite{toappear}, authors employ the scramblon effective theory to study wormhole teleportation using the linear response theory. In this work, we go beyond this perturbative regime. By summing up all diagrams with different scramblon configurations, we investigate the fidelity of wormhole teleportation. This involves computing both mutual information and entanglement negativity, including $1/N$ corrections that are crucial in realistic experiments. 
    
    Following conventions in \cite{toappear}, we distinguish between two distinct parameter regimes for wormhole teleportation. The regime characterized by strong coupling ($\mu\sim \mathcal{O}(1)$) between two sides is termed the probe limit, holding particular significance for simulating semi-classical gravity. In this regime, our results address the crucial question of determining the minimum value of $N$ necessary to observe sharp signatures of the wormhole through quantum simulations of the SYK model with $N$ Majorana fermions. A naive estimate might suggest that $N\sim40$ is sufficient, for which state-of-the-art numerics for OTOCs show good agreement with theoretical analysis \cite{Kobrin:2020xms}. However, as we will elaborate, this turns out to be insufficient. Achieving a semi-classical limit necessitates both $\mu N\gg 1$ and $\mu \ll 1$, requiring a significantly larger value for $N\sim 10^3$ in practice. On the other hand, when the coupling is weak ($\mu\sim \mathcal{O}(1/N)$), finite-size effects become inevitable even for arbitrarily large $N$, corresponding to quantum gravity with strong backreactions. In addition to the above discussions for strongly interacting systems with near-maximal chaos, we also apply our general formula to weakly interacting (or high-temperature) systems for comparison.

    This paper is organized as follows: In Section \ref{sec:setup}, we elaborate on the setup of the wormhole teleportation protocol and provide a brief introduction to the scramblon effective theory. Following that, we present the general framework for computing matrix elements of the reduced density matrix in Section \ref{sec:maincalculation}. In Section \ref{sec:strong}, we analyze the results with strong couplings, using the example of the large-$q$ SYK model and focusing on finite $N$ corrections. We also discuss teleportation in the weak coupling case.

\begin{figure}[t]
    \centering
   
    \subfloat[Traversable wormhole]{
	\includegraphics[width=0.36\linewidth]{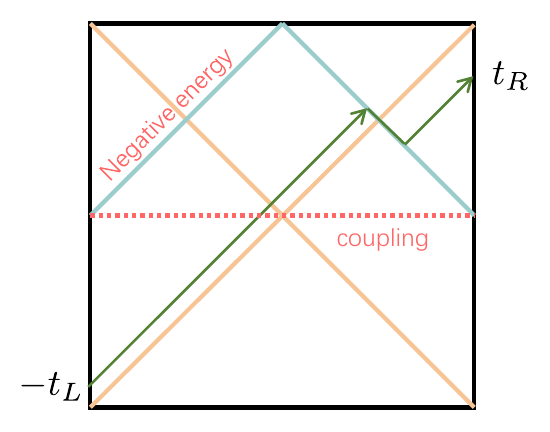}
}
\hspace{5pt}
\subfloat[Teleportation protocol]{
	\includegraphics[width=0.30\linewidth]{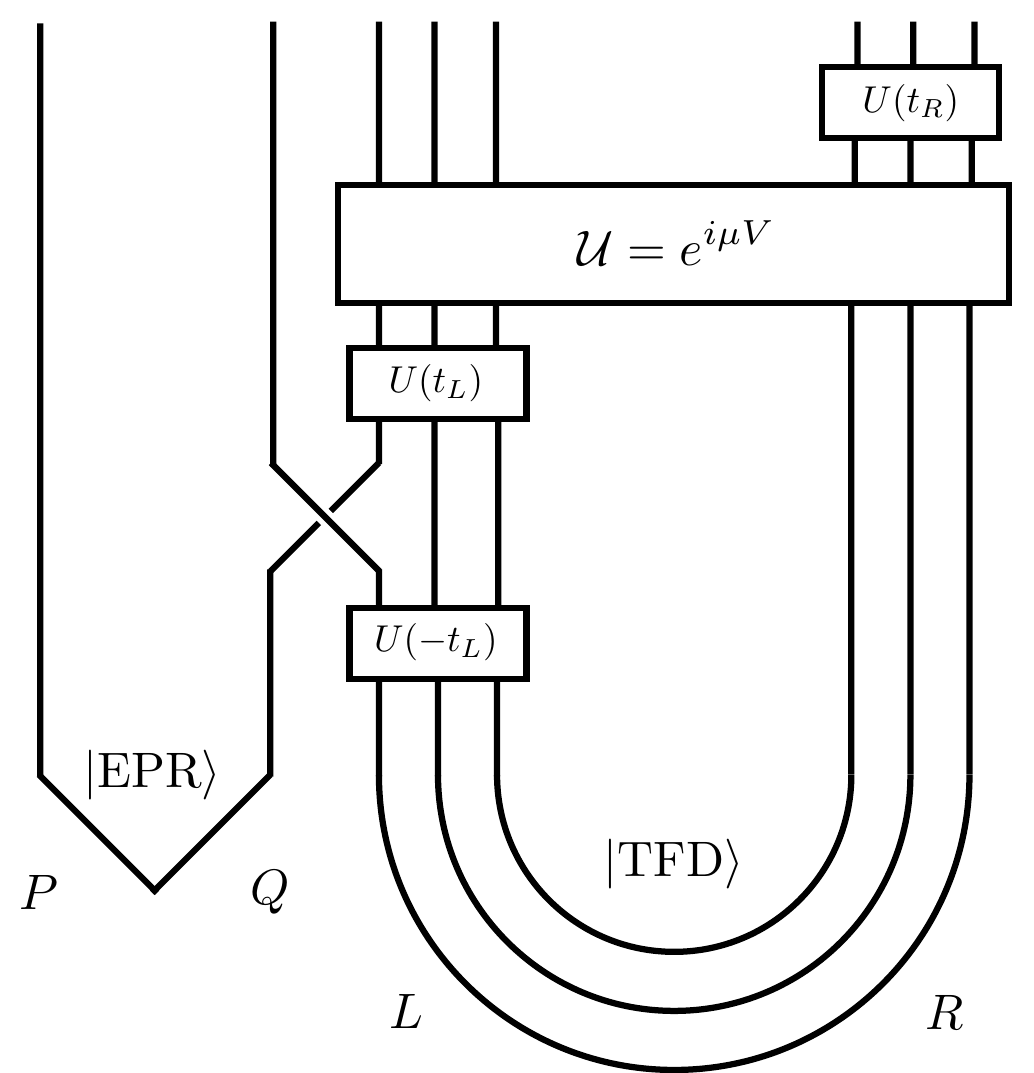}
}
    \caption{We present illustrations of (a). Traversable wormhole via double trace deformation on the gravity side and (b). Wormhole teleportation protocol motivated by the gravity picture.}
    \label{fig: protocol}
    \end{figure}

\section{Setup}\label{sec:setup}

 In this section, we elaborate the setup of the wormhole teleportation protocol, with a specific emphasis on the teleportation of a single qubit. Subsequently, we explain the close relationship between teleportation and information scrambling in chaotic systems, highlighting the efficacy of the scramblon effective theory.
 \subsection{Wormhole teleportation protocol}
Without violating the averaged null energy \cite{Graham:2007va,Kelly:2014mra,Kontou:2012ve,Kontou:2015yha,Wall:2009wi}, wormholes are shown to be non-traversable \cite{Hochberg:1998ii,Morris:1988tu,Visser:1995cc,Visser:2003yf}. A concrete method to overcome this obstacle was proposed in a seminal paper by Gao, Jafferis, and Wall \cite{Gao:2016bin} through coupling two boundaries via a double-trace deformation. The original discussion is based on a two-sided black hole geometry in asymptotically anti-de Sitter (AdS) spacetime. A message, represented by some particle, is created on the left boundary at a very early time $-t_L$. Without any further operation, it will fall into the black hole following the geodesic and ultimately be destroyed by the singularity. Now, we attempt to rescue the message by applying instantaneous coupling between the two sides, represented by $\mathcal{U}=e^{i\mu \phi_L \phi_R}$ at $t=0$, where $\phi_{L}$ and $\phi_{R}$ are boundary operators corresponding to certain bulk quantum fields. Such a perturbation injects energy into the system, which can be made negative for a specific sign of $\mu$. Therefore, the message is scattered by the negative energy current before it reaches the singularity. The shock wave calculation for this scattering problem \cite{Dray:1984ha,Dray:1985yt} predicts that the message then experiences a time advance, and is able to be received by an observer on the right boundary, as illustrated in FIG. \ref{fig: protocol} (a). This demonstrates the traversability of the wormhole. 

    Motivated by this development, a wormhole teleportation protocol \cite{Brown:2019hmk,Nezami:2021yaq,Maldacena:2017axo,Gao:2018yzk,Gao:2019nyj,Schuster:2021uvg} is proposed to simulate traversable wormholes in quantum many-body systems with high controllability. The corresponding quantum circuit is shown in FIG. \ref{fig: protocol} (b). It operates on a quantum many-body system with $2+N$ qubits, with the message $Q$ described by a single qubit, its reference $P$, and two copies of some chaotic quantum many-body system each containing $N/2$ qubits. The total Hamiltonian of the many-body system is the sum of the left and right Hamiltonians: $H_\text{tot}=H_L+H_R$ where $H_L=H\otimes I$ and $H_R=I\otimes H^T$. Qubits are initialized in $|\psi_0\rangle=|\text{EPR}\rangle_{PQ}\otimes |\text{TFD}\rangle_{LR}$. Here, $|\text{EPR}\rangle_{PQ}=(|0\rangle_P\otimes |0\rangle_Q+|1\rangle_P\otimes |1\rangle_Q)/\sqrt{2}$ is the Einstein-Podolsky-Rosen (EPR) state and $|\text{TFD}\rangle_{LR}$ is the thermofield double (TFD) state \cite{Israel:1976ur,Maldacena:2001kr} defined as
\begin{equation}
    |\text{TFD}\rangle_{LR}=\frac{1}{\sqrt{Z}}\sum_ne^{-\beta E_n/2}|E_n\rangle_L\otimes \overline{|E_n\rangle}_R,
    \end{equation}
    where $|E_n\rangle$ represents an eigenstate of the Hamiltonian $H$ with eigenenergy $E_n$, and the bar denotes the complex conjugate in the computational basis. $Z=\text{tr}[e^{-\beta H}]$ is a normalization factor that matches the thermal partition function. As in the gravity setup, we first send our signal $Q$ into the left system at $t=-t_L$. This is achieved by evolving the entire system backwards to $-t_L$ under $H_\text{tot}$ and performing a swap operation between $Q$ and the first qubit in $L$, denoted as $L_1$. The generalization of this protocol to multi-qubit messages is straightforward. The system is then evolved to $t=0$, when an instantaneous coupling $\mathcal{U}=e^{i\mu\sum_{i=1}^N O^i_L O^i_R}$ is applied to the system for a set of operators $\{(O^i_L,O^i_R)\}$. Finally, we ask whether the message appears on $R_1$, the first qubit in $R$, at time $t=t_R$ by computing the reduced density matrix $\rho_{PR_1}$. 

To be concrete, we focus on quantum many-body systems described by $N$ Majorana fermions $\psi_{L,R}^{k}$ with all-to-all interactions, such as the SYK model \cite{Kit.KITP.2,Maldacena:2016hyu} and its various generalizations \cite{Chowdhury:2021qpy,Zhang:2022yaw}. The canonical commutation relation reads $\{\psi^{j}_\alpha,\psi^{k}_\beta\}=\delta^{jk}\delta_{\alpha\beta}$. In particular, $L_1$ or $R_1$ contains Majorana fermions $\psi^{1/2}_L$ or $\psi^{1/2}_R$ respectively. Similarly, we represent each qubit in $P$ or $Q$ by two Majorana fermions, denoted as $\psi^{1/2}_P$ or $\psi^{1/2}_Q$. Additionally, we adopt the convention that the EPR state \cite{Gu:2017njx,Maldacena:2018lmt} corresponds to the vacuum of two complex fermions $c^k=\psi^k_P+i\psi^k_Q$ with $k=1,2$, satisfying $c^k|\text{EPR}\rangle_{PQ}=0$. The reduced density matrix $\rho_{PR_1}$ can be expanded in an orthonormal basis made up of products of Majorana fermions, which reads
    \begin{equation}
    \begin{aligned}
    \rho_{PR_1}=&\frac{1}{4}\Big( I+i\rho_{1}^{(2)}\psi_{P}^{1}\psi_{P}^{2}+i\rho_{2}^{(2)}\psi_{R}^{1}\psi_{R}^{2}+i\sum_{j,k}\rho_{j,k}^{(2)}\psi_{P}^{j}\psi_{R}^{k}+\rho^{(4)}\psi_{P}^{1}\psi_{P}^{2}\psi_{R}^{1}\psi_{R}^{2} \Big).
    \end{aligned}
    \end{equation} 
    This expression incorporates all possible terms with even fermion parity. Assuming the Hamiltonian $H$ is invariant under O(2) rotations between first two Majorana modes further add restrictions that $\rho_{1,1}^{(2)}=\rho_{2,2}^{(2)}\equiv\rho^{(2)}$ and $\rho_{1}^{(2)}=\rho_{2}^{(2)}=\rho_{1,2}^{(2)}=\rho_{2,1}^{(2)}=0$. It is then straightforward to express $\rho^{(2)}$ and $\rho^{(4)}$ as \cite{Gao:2019nyj}
    \begin{equation}\label{eqn:real_is_important}
    \rho^{(2)}=2\text{Re}[\mathcal{I}_3-2\mathcal{I}_4] ,\ \ \ \ \ \ \rho^{(4)}=8\text{Re}[\mathcal{I}_2-\mathcal{I}_1].
    \end{equation}
    Here, we have introduced correlation functions defined only on $L\cup R$
    \begin{equation}\label{eqn:correlator}
    \begin{aligned}
    \mathcal{I}_1&=\langle \psi_L^1\psi_L^2 \mathcal{U}^\dagger \psi_R^1\psi_R^2 \mathcal{U}\rangle,\ \ \ \ \ \ \ \mathcal{I}_2=\langle \psi_L^1 \mathcal{U}^\dagger \psi_R^1\psi_R^2 \mathcal{U}\psi_L^2\rangle,\\
    \mathcal{I}_3&=\langle \psi_L^1 \mathcal{U}^\dagger \psi_R^1 \mathcal{U}\rangle, \ \ \ \ \ \ \ \ \ \ \ \ \ \ \mathcal{I}_4=\langle \psi_L^1\psi_L^2 \mathcal{U}^\dagger \psi_R^1 \mathcal{U}\psi_L^2\rangle.
    \end{aligned}
    \end{equation}
    Here, the expectation is computed with respect to the TFD state $|\text{TFD}\rangle_{LR}$. For conciseness, we have omitted the time arguments for all operators, since, all left Majorana operators have time $-t_L$ and all right Majorana operators have time $t_R$. The coupling $\mathcal{U}$ and $\mathcal{U}^\dagger$ are inserted at $t=0$. After computing the reduced density matrix $\rho_{PR_1}$, we can investigate the correlation/entanglement between $P$ and $R_1$ by computing mutual information and entanglement negativity. 

\subsection{Teleportation and scramblons}\label{subsec:scramblon}
    We are mainly interested in the teleportation signal with small $\mu\ll 1$. This emphasis stems from two key reasons. Firstly, the teleportation only becomes non-trivial when $\mu\ll 1$, Otherwise, we are asking whether we can transmit some quantum information after a significant coupling between two sides, which always yields a positive answer. Secondly, and more importantly, $\mu\ll 1$ is crucial for a justified simulation of traversable wormholes: All quantum mechanical models with finite local Hilbert space dimensions are UV complete. Consequently, semiclassical gravity can only emerge within specific sectors, typically in the low-energy limit. The coupling serves as a perturbation to the system. When $\mu$ is large, it effectively drives the system out of the low-energy manifold, rendering the holographic duality no longer valid.

Naively, a small $\mu$ allows for a perturbative calculation of correlation functions. However, the expansion in $\mu$ ultimately becomes problematic for large $t_L$ and $t_R$ due to quantum many-body chaos \cite{Maldacena:2015waa,Shenker:2013pqa,Roberts:2014isa,Shenker:2014cwa,kitaev2014talk}. As an example, expanding $\mathcal{I}_3$ in $\mu$ gives
    \begin{equation}
    \mathcal{I}_3\approx \langle \psi_L^1 \psi_R^1 \rangle+i\mu\sum_i\Big(\langle \psi_L^1 \psi_R^1 O^i_L O^i_R\rangle-\langle \psi_L^1  O^i_L O^i_R\psi_R^1\rangle\Big).
    \end{equation}
    Using $I\otimes O~|\text{TFD}\rangle= e^{-\beta H/2}O^Te^{\beta H/2}\otimes I~|\text{TFD}\rangle$ for arbitrary operator $O$, one can check that the second term in the bracket is an OTOC since
    \begin{equation}
    \langle \psi_L^1  O_L O'_R\psi_R^1\rangle=i{Z^{-1}}\text{tr}_L[\psi^1(-t_L)Oe^{-\frac{\beta H}{2}}\psi^1(-t_R)O'^Te^{-\frac{\beta H}{2}}].
    \end{equation}
    Here, we have assumed that the operator $O$ is bosonic, and generalizations to the fermionic case are straightforward. In chaotic many-body systems, we therefore expect an exponential deviation behavior of OTOC due to information scrambling, yielding $\mathcal{I}_3 \approx \langle \psi_L^1 \psi_R^1 \rangle \left(1 - c \mu e^{\varkappa (t_L + t_R)/2}\right)$, where $\varkappa$ denotes the quantum Lyapunov exponent, and $c$ is a complex constant. As a consequence, it is necessary to perform a resummation over $\mu e^{\varkappa (t_L + t_R)/2}$ to describe the behavior of the teleportation signal. 

Scramblon effective theory provides an efficient way to sum up all exponential growing contributions in correlators with OTO correlations \cite{Gu:2021xaj}. It visualizes the exponential growth as the propagator of some collective mode, named scramblon \cite{Gu:2021xaj,Kitaev:2017awl,Gu:2018jsv}. For $\varkappa t\gg 1$, the contribution from all other modes is negligible compared to the scramblon mode, and operators only interact through exchanging scramblons. This is compatible with our requirement of $\mu \ll 1$, for which a significant teleportation signal only appears when $\varkappa t \sim \ln \mu^{-1}$. As an illustration for the scramblon calculation, let us consider the four-point function:
    \begin{equation}
    F(\{\theta_i\})=\langle T_\tau \psi^1(-i\theta_1)\psi^1(-i\theta_2)\psi^2(-i\theta_3)\psi^2(-i\theta_4) \rangle_\beta,
    \end{equation}
    where $\theta_i = \tau_i + it_i$ denotes complex time, and $T_\tau$ represents the imaginary time ordering operator. We focus on $t_1\approx t_2 \gg t_3\approx t_4$ and $\tau_1\geq \tau_3\geq \tau_2\geq \tau_4\geq \tau_1-\beta$, for which $F(\{\theta_i\})$ becomes an OTOC. In the scramblon picture, the pair of $\psi^2$ operators in the past can emit an arbitrary number of scramblons, which propagate freely and are subsequently absorbed by the pair of $\psi^1$ operators in the future. This gives 
    \begin{equation}
    \begin{aligned}\label{eqn:OTOC}
    F(\{\theta_i\})=&\begin{tikzpicture}[scale=1.15]
    \node[bvertexnormal] (R) at (-20pt,0pt) {};
    \node (Ra) at (-20pt,0pt) {};
    \draw[thick] (R) -- ++(135:18pt) node[left]{\scriptsize${\theta_1}$};
    \draw[thick] (R) -- ++(-135:18pt) node[left]{\scriptsize${\theta_2}$};
    \node[bvertexnormal] (A) at (20pt,0pt) {};
    \node (Aa) at (20pt,0pt) {};
    \draw[thick] (A) -- ++(45:18pt) node[right]{\scriptsize${\theta_3}$};
    \draw[thick] (A) -- ++(-45:18pt) node[right]{\scriptsize${\theta_4}$};
    \draw[wavy] (R) to[out=30,in=150] (A);
    \draw[wavy] (R) to[out=0,in=180] (A);
    \draw[wavy] (R) to[out=-30,in=210] (A);
    \end{tikzpicture} =\sum_{m=0}^{\infty}\frac{(-\lambda)^{m}}{m!}\Upsilon^{\text{R},m}(\theta_{12})\Upsilon^{\text{A},m}(\theta_{34}),
    \end{aligned}
    \end{equation}
    where we use solid/wavy lines to present the propagation of Majorana fermions/scramblons. $\lambda=C^{-1}e^{i\varkappa(\beta/2-\theta_{1}-\theta_{2}+\theta_{3}+\theta_{4})/2}$ refers to the scramblon propagator, with $C$ as a constant of order $N$. $m!$ appears as the symmetry factor of the diagram. $\Upsilon^{\text{R},m}$ and $\Upsilon^{\text{A},m}$ represent scattering vertices between Majorana fermions and $m$ scramblons in the future and the past, respectively. They have been computed in closed-form to the zeroth order in $1/N$ in the large-$q$ SYK model \cite{Gu:2021xaj}, Brownian SYK models \cite{Gu:2021xaj,Zhang:2023vpm,Zhang:2022knu}, and Brownian circuits \cite{Liu:2023lyu}, where we further have constraints $\Upsilon^{\text{R},m}=\Upsilon^{\text{A},m}$ due to the time-reversal symmetry. For $m=0$, both scattering vertices reduce to the imaginary-time two-point function $G(\theta)=\langle e^{\theta H}\psi^1 e^{-\theta H}\psi^1\rangle_\beta$.

    Besides, we introduce an alternative picture for OTOC, which will prove beneficial in subsequent analyses. We define generating functions for the scattering vertices:
    \begin{equation}
f^{\text{R}/\text{A}}(x,\theta_{12})=
    \begin{tikzpicture}[scale=1.15]
    \node[bvertexnormal] (R) at (-30pt,0pt) {};
\node[sdot] (D1) at (-5pt,12pt) {};
\node[sdot] (D2) at (-5pt,0pt) {};
\node[sdot] (D3) at (-5pt,-12pt) {};
\node at (0pt,12pt) {\scriptsize $x$};
\node at (0pt,0pt) {\scriptsize $x$};
\node at (0pt,-12pt) {\scriptsize $x$};
\draw[thick] (R) -- ++(135:20pt) node[left]{\scriptsize$\theta_1$};
\draw[thick] (R) -- ++(-135:20pt) node[left]{\scriptsize$\theta_2$};
\draw[wavy] (D1) to[out=180,in=40] (R);
\draw[wavy] (D2) to (R);
\draw[wavy] (D3) to[out=-180,in=-40] (R);
    \end{tikzpicture}=\sum_{m=0}^{\infty}\frac{(-x)^{m}}{m!}\Upsilon^{\text{R}/\text{A},m}(\theta_{12}).
    \end{equation}
    The functions $f^{\text{R}/\text{A}}(x,\theta_{12})$ can be interpreted as two-point functions with a condensation of scramblon fields since we replaced each scramblon line with a constant $-x$. This offers a means to probe the strength of the scramblon field. We further express $f^{\text{R}/\text{A}}(x,\theta_{12})$ as a Laplace transform
    \begin{equation}\label{eqn:h_def}
    f^{\text{R}/\text{A}}(x,\theta)=\int_0^\infty dy~h^{\text{R}/\text{A}}(y,\theta)\exp({-xy}).
    \end{equation}
    Here, $y$ represents the perturbation strength resulting from the insertion of the Majorana fermion operator with a distribution function $h^{\text{R}/\text{A}}(y,\theta_{12})$. This strength couples to the scramblon field through a simple action $S=xy$, analogous to shock waves \cite{Dray:1984ha,Dray:1985yt}. By expanding \eqref{eqn:h_def} in $x$, we can identify $\Upsilon^{\text{R/A},m}(\theta)=\int_0^\infty dy~y^m h^{\text{R}/\text{A}}(y,\theta)$. Therefore, Eq. \eqref{eqn:OTOC} can be equivalently written as
    \begin{equation}\label{eqn:OTOChf}
    F(\{\theta_i\})=\int_0^\infty dy~f^\text{R}(\lambda y,\theta_{12})h^\text{A}( y,\theta_{34}).
    \end{equation}
    Therefore, OTOC describes a measurement of perturbations induced by $\psi^2$ operators through computing the two-point function of $\psi^1$. The perturbation is amplified by $\lambda$, the propagation of scramblons, due to quantum many-body chaos. As we will see, a similar picture emerges when computing $\mathcal{I}_k$. 

\begin{figure}[t!]
    \centering
    \includegraphics[width=0.5\linewidth]{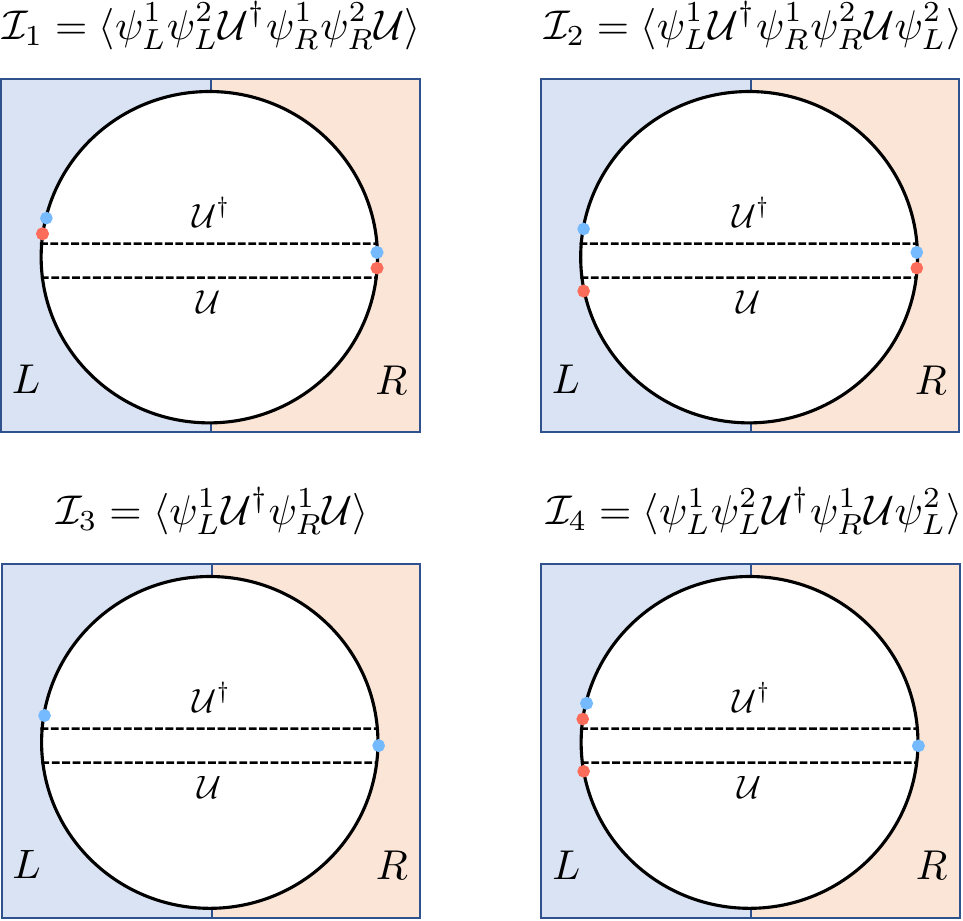}
    \caption{We depict the operator insertions in various correlation functions $\mathcal{I}_k$. In this visualization, each black half-circle signifies the TFD state, while blue and red dots correspond to $\psi^1$ and $\psi^2$ operators, respectively. All left operators are inserted at $t=-t_L$, and all right operators are inserted at $t=t_R$. The dashed lines represent the coupling between the two sides. }
    \label{fig: configuration}
    \end{figure}

\section{Calculation of the Density Matrix}\label{sec:maincalculation}
    In this section, we derive the result of $\rho_{PR_1}$ by calculating $\mathcal{I}_k$ using the scramblon effective theory. In \cite{toappear}, it is explained that the teleportation manifests two distinct regimes contingent on the scaling of the coupling strength $\mu$ with $N$, each dominated by a distinct group of diagrams. Nevertheless, we anticipate that this disparity becomes less significant for finite $N$ systems. Consequently, our approach involves summing all diagrams with arbitrary scramblon configurations.

    We first discuss $\mathcal{I}_3=\langle\psi_L^1 \mathcal{U}^\dagger \psi_R^1 \mathcal{U}\rangle$, which exclusively involves a single pair of Majorana operators. This is directly related to the linear response function. Instead of executing a swap operation, we can represent the message by introducing a perturbation $e^{i\eta \psi^1_L}$ \footnote{Here, $\eta$ is some Grassmann number.}. The expectation $\langle \psi^1_R (t_R)\rangle$, to the leading order in $\eta$, is determined by the celebrated Kubo formula \cite{altland2010condensed}, yielding $\langle \psi^1_R (t_R)\rangle \approx -i\eta~\text{Re}[\mathcal{I}_3]$. This linear response problem has been thoroughly investigated in \cite{toappear}, where $\mathcal{I}_3$ has been computed in closed-form. Let us take the coupling between two sides as $\mathcal{U}=e^{-\mu \sum_{i=1}^N \psi^i_L\psi^i_R }$. The result reads
    \begin{equation}\label{eqn:I3}
    \begin{aligned}
    \mathcal{I}_3&=\langle\psi_L^1 \mathcal{U}^\dagger \psi_R^1 \rangle\langle\mathcal{U}\rangle=\langle\mathcal{U}\rangle
    \begin{tikzpicture}[scale=1.15]
    \node[bvertexnormal] (R) at (-30pt,0pt) {};
    \node[sdot] (D1) at (-5pt,12pt) {};
    \node[sdot] (D2) at (-5pt,0pt) {};
    \node[sdot] (D3) at (-5pt,-12pt) {};
    \node at (2pt,15pt) {};
    \node at (2pt,0pt) {};
    \node at (2pt,-15pt) {};
    \draw[thick] (R) -- ++(135:20pt) node[left]{ };
    \draw[thick] (R) -- ++(-135:20pt) node[left]{ };
    \draw[wavy] (D1) to[out=160,in=45] (R);
    \draw[wavy] (D1) to[out=240,in=35] (R);
    \draw[wavy] (D2) to (R);
    \draw[wavy] (D3) to[out=-180,in=-40] (R);
    \end{tikzpicture} \\
    &=i e^{-i\mu N G(\frac{\beta}{2})}\int_0^\infty dy~ h^\text{A}\Big(y,\frac{\beta}{2}+it_{LR}\Big)e^{i\mu Nf^\text{R}(e^{-i\varkappa \beta/4}\lambda_0 y,\frac{\beta}{2})}.
    \end{aligned}
    \end{equation}
    Here, we have introduced $t_{LR}=t_L-t_R$ and the norm of the scramblon propagator $\lambda_0=e^{\varkappa (t_L+t_R)/2}/C$. Generalizations to general $O^i_L$ and $O^i_R$ only require replacing $G$ and $f^\text{R}$ with their counterparts for $O$. In the first line, we observe that $\psi^1$ operators are only out-of-time ordered with respect to $\mathcal{U}^\dagger$, as shown in FIG. \ref{fig: configuration}. Therefore, $\mathcal{U}$ does not interact with $\psi^1$ in the scramblon effective theory. Summing up all diagrams includes: (i) expanding $\mathcal{U}^\dagger$ to $m$-th order in $\mu$ (which creates $m$ source terms represented by black dots), and (ii) each source emitting $n_j$ scramblons, leading to the final result. Compared to \eqref{eqn:OTOChf}, the only distinction lies in the exponentialization of the probe function $f^\text{R}$ in $\mathcal{I}_3$. It's worth noting that this calculation presents a close analogy to the 2D Jackiw-Teitelboim gravity, showcasing the potential for an emergent holographic framework concerning wormhole teleportation in chaotic systems with all-to-all interactions.

    \subsection{Relating \texorpdfstring{$\mathcal{I}_1$}{TEXT} and \texorpdfstring{$\mathcal{I}_2$}{TEXT} to \texorpdfstring{$\mathcal{I}_3$}{TEXT}}
    With this understanding, we proceed to compute $\mathcal{I}_1$ and $\mathcal{I}_2$, which turn out to be closely related to $\mathcal{I}_3$. Let us begin by analyzing $\mathcal{I}_1$. As illustrated in FIG. \ref{fig: configuration}, $\mathcal{I}_1$ takes a similar form to $\mathcal{I}_3$: $\mathcal{U}$ does not exhibit any OTO correlation with either $\psi^1$ or $\psi^2$. Following the derivation giving Eq. \eqref{eqn:I3}, we anticipate
    \begin{equation}\label{eqn:I1_1}
    \begin{aligned}
    \mathcal{I}_1
    &=e^{-i\mu N G(\frac{\beta}{2})}\int_0^\infty dy~ h^\text{A}_{\psi^1\psi^2}\Big(y,\frac{\beta}{2}+it_{LR}\Big)e^{i\mu Nf^\text{R}(e^{-i\varkappa \beta/4}\lambda_0 y,\frac{\beta}{2})}.
    \end{aligned}
    \end{equation}
    Up to an overall phase factor, the only distinction is to replace the perturbation distribution function of $\psi^1$, $h^\text{A}(y,\theta)$, with the distribution $h^\text{A}_{\psi^1\psi^2}(y,\theta)$ for the composite operator $\psi^1\psi^2$. The remaining task is to find the expression for $h^\text{A}_{\psi^1\psi^2}(y,\theta)$. This turns out to be simple, within the assumption of $\langle \psi^1\psi^2\rangle=0$, for which $\psi^1$ and $\psi^2$ interact independently with scramblons. Diagrammatically, we have
    \begin{equation}\label{eqn:composite}
    \begin{aligned}
    \begin{tikzpicture}[scale=1.15]
    \node[bvertexnormal] (R) at (-30pt,0pt) {};
    \node(R1) at (-30pt,1pt) {};
    \node (R2) at (-30pt,-1pt) {};
    \node (D1) at (-8pt,12pt) {};
    \node (D2) at (-8pt,0pt) {};
    \node (D3) at (-8pt,-12pt) {};
    \node at (0pt,12pt) {};
    \node at (0pt,0pt) {};
    \node at (0pt,-12pt) {};
    \draw[thick,red] (R1) -- ++(135:20pt) node[left]{};
    \draw[thick,red] (R1) -- ++(-135:20pt) node[left]{};
    \draw[thick,blue] (R2) -- ++(135:20pt) node[left]{};
    \draw[thick,blue] (R2) -- ++(-135:20pt) node[left]{};
    \draw[wavy] (D1) to[out=-180,in=40] (R);
    \draw[wavy] (D2) to (R);
    \draw[wavy] (D3) to[out=-180,in=-40] (R);
    \node[bvertexnormal] (R) at (-30pt,0pt) {};
    \end{tikzpicture} \hspace{-10pt}
    &=\sum_{m=0}^n C_n^m
    \begin{tikzpicture}[scale=1.15]
    \node (R2) at (-30pt,-1pt) {};
    \node (D1) at (-8pt,12pt) {};
    \node at (0pt,12pt) {};
    \node at (0pt,0pt) {};
    \node at (0pt,-12pt) {};
    \draw[thick,blue] (R2) -- ++(135:20pt) node[left]{};
    \draw[thick,blue] (R2) -- ++(-135:20pt) node[left]{};
    \draw[wavy] (D1) to[out=-180,in=40] (R);
    \node[bvertexnormal] (R2) at (-30pt,0pt) {};
    \end{tikzpicture}\hspace{-10pt}\times
    \begin{tikzpicture}[scale=1.15]
    \node(R1) at (-30pt,1pt) {};
    \node (D2) at (-8pt,0pt) {};
    \node (D3) at (-8pt,-12pt) {};
    \node at (0pt,12pt) {};
    \node at (0pt,0pt) {};
    \node at (0pt,-12pt) {};
    \draw[thick,red] (R1) -- ++(135:20pt) node[left]{};
    \draw[thick,red] (R1) -- ++(-135:20pt) node[left]{};
    \draw[wavy] (D2) to (R);
    \draw[wavy] (D3) to[out=-180,in=-40] (R);
    \node[bvertexnormal] (R1) at (-30pt,0pt) {};
    \end{tikzpicture}\\
    \Upsilon^{R/A,n}_{\psi^1\psi^2}(\theta)&=\sum_{m=0}^n C_n^m\Upsilon^{R/A,m}(\theta)\Upsilon^{R/A,n-m}(\theta).
    \end{aligned}
    \end{equation}
    Here, red and blue lines correspond to $\psi^1$ and $\psi^2$ operators, respectively. For a scattering vertex between a pair of $\psi^1\psi^2$ operators and $n$ scramblons, we sum up contributions where $m$ scramblons are scattered by $\psi^1$ and $n-m$ scramblons are scattered by $\psi^2$. $C^m_n$ is the binomial number that appears in counting the number of equivalent diagrams. This leads to $f^{R/A}_{\psi^1\psi^2}(x,\theta)=f^{R/A}(x,\theta)^2$ and thus 
    \begin{equation}\label{eqn:I1_2}
    \begin{aligned}
    h^{R/A}_{\psi^1\psi^2}(y,\theta)=\int_0^y dy'~h^{R/A}(y',\theta)h^{R/A}(y-y',\theta),
    \end{aligned}
    \end{equation}
    due to the convolution theorem. Combining Eq. \eqref{eqn:I1_1} and \eqref{eqn:I1_2}, we obtain the result of $\mathcal{I}_1$.  

    Next, we turn our attention to the computation of $\mathcal{I}_2$. According to FIG. \ref{fig: configuration}, both $\mathcal{U}$ and $\mathcal{U}^\dagger$ show non-trivial OTO correlation. Nevertheless, their contribution factorizes: through exchanging scramblons, $\mathcal{U}$ only interacts with $\psi^2$ operators and $\mathcal{U}^\dagger$ only interacts with $\psi^1$ operators. Therefore, we have 
    \begin{equation}\mathcal{I}_2=
    \begin{tikzpicture}[scale=1.15]
    \node[bvertexnormal] (R) at (-30pt,0pt) {};
    \node[sdot] (D1) at (-5pt,12pt) {};
    \node[sdot] (D2) at (-5pt,0pt) {};
    \node[sdot] (D3) at (-5pt,-12pt) {};
    \node at (2pt,15pt) {};
    \node at (2pt,0pt) {};
    \node at (2pt,-15pt) {};
    \draw[thick,blue] (R) -- ++(135:20pt) node[left]{ };
    \draw[thick,blue] (R) -- ++(-135:20pt) node[left]{ };
    \draw[wavy] (D1) to[out=160,in=45] (R);
    \draw[wavy] (D1) to[out=240,in=35] (R);
    \draw[wavy] (D2) to (R);
    \draw[wavy] (D3) to[out=-180,in=-40] (R);
    \end{tikzpicture} \times
    \begin{tikzpicture}[scale=1.15]
    \node[bvertexnormal] (R) at (-30pt,0pt) {};
    \node[sdot,brown] (D1) at (-5pt,12pt) {};
    \node[sdot,brown] (D2) at (-5pt,0pt) {};
    \node[sdot,brown] (D3) at (-5pt,-12pt) {};
    \node at (2pt,15pt) {};
    \node at (2pt,0pt) {};
    \node at (2pt,-15pt) {};
    \draw[thick,red] (R) -- ++(135:20pt) node[left]{ };
    \draw[thick,red] (R) -- ++(-135:20pt) node[left]{ };
        \draw[wavy] (D3) to[out=-160,in=-45] (R);
    \draw[wavy] (D3) to[out=-240,in=-35] (R);
    \draw[wavy] (D2) to (R);
    \draw[wavy] (D1) to[out=180,in=40] (R);
    \end{tikzpicture}=|\mathcal{I}_3|^2,
    \end{equation}
    where the brown dots represent sources originating from the expansion of $\mathcal{U}$. The summation over scramblon configurations in each diagram is independent, thus leading to factorization of the final result.

 \subsection{Calculation of \texorpdfstring{$\mathcal{I}_4$}{TEXT}}
    Finally, we delve into the computation of $\mathcal{I}_4$, which involves summing up a new set of diagrams. From FIG. \ref{fig: configuration}, it is evident that both $\mathcal{U}$ and $\mathcal{U}^\dagger$ play a non-trivial role. While the pair of $\psi^1$ operators only exhibits OTO correlation with $\mathcal{U}^\dagger$, the $\psi^2$ operators show OTO correlations with both $\mathcal{U}^\dagger$ and $\mathcal{U}$. Consequently, the relevant scramblon diagrams take the following form:
\begin{equation}\label{eqn:diaI4}
\mathcal{I}_4=
    \begin{tikzpicture}[scale=1.15]
    \node[bvertexnormal] (R) at (-30pt,0pt) {};
    \node[bvertexnormal] (A) at (30pt,0pt) {};
    \node[sdot] (D1) at (-0pt,12pt) {};
    \node[sdot] (D2) at (-0pt,0pt) {};
    \node[sdot,brown] (D3) at (-0pt,-12pt) {};
    \draw[thick,blue] (R) -- ++(135:20pt) node[left]{ };
    \draw[thick,blue] (R) -- ++(-135:20pt) node[left]{ };
    \draw[wavy] (D1) to[out=160,in=45] (R);
    \draw[wavy] (D1) to[out=240,in=35] (R);
    \draw[wavy] (D2) to (R);
    \draw[wavy] (D3) to[out=0,in=220] (A);
    \draw[wavy] (D2) to[out=-45,in=190] (A);
    \draw[wavy] (D2) to[out=45,in=170] (A);

    \draw[thick,red] (A) -- ++(45:20pt) node[right]{ };
    \draw[thick,red] (A) -- ++(-45:20pt) node[right]{ };
    \end{tikzpicture}
    \end{equation}
    It describes the process in which each black dot can emit an arbitrary number of scramblons, which are absorbed by either $\psi^1$ or $\psi^2$, and all scramblons emitted by brown dots are absorbed by $\psi^2$. This leads to the expression
    
    \begin{equation}
    \begin{aligned}
    \mathcal{I}_4=&{-i}\sum_p\sum_q\frac{(i\mu{N})^{p}(-i\mu{N})^{q}}{p!q!}\sum_{n_1^1,...,n_p^1}\sum_{n_1^2,...,n_p^2}\sum_{m_1,...,m_q}\Upsilon^{\text{A},\sum_k n_k^1}\Upsilon^{\text{A},\sum_k n_k^2+\sum_b m_b}\\
    &\times\prod_j\frac{(-\lambda_0e^{-i\frac{\varkappa \beta}{4}})^{n_j^1}({-\lambda_{1}})^{n_j^2}}{ n_j^1!n_j^2!}\Upsilon^{\text{R},n^1_j+n^2_j}\prod_a\frac{({-\lambda_{1}})^{m_a}}{ m_a!}\Upsilon^{\text{R},m_a},
    \end{aligned}
    \end{equation}
    with $\lambda_{1}=e^{\varkappa t_{L}}/C$. Here, we omit all arguments of $\Upsilon^{\text{R/A}}$ for conciseness. The variables $p$ and $q$ denote the number of black dots and brown dots in \eqref{eqn:diaI4}, respectively. Each black dot $j \in {1,2,...,p}$ emits $n_j^1 + n_j^2$ scramblons, where $n_j^1$ of them interact with the blue fermion lines, and $n_j^2$ interact with the red fermion lines. Each brown dot $a \in {1,2,...,q}$ emits $m_a$ scramblons, which only interact with the red fermions. Different phases of the scramblon propagator date back to different imaginary time configurations of operators. By expressing all vertex functions using $h^{\text{R/A}}$, we find
    \begin{equation}\label{eqn:I4}
    \begin{aligned}
    \mathcal{I}_4=-i&\int_0^\infty dy_1dy_2~h^\text{A}\Big(y_1,\frac{\beta}{2}+it_{LR}\Big)h^\text{A}\Big(y_2,0\Big) e^{i\mu N \bigg[f^\text{R}\Big(\lambda_0 y_1e^{-i\frac{\varkappa \beta}{4}}+{\lambda_1}y_2,\frac{\beta}{2}\Big)-f^\text{R}\Big({\lambda_1}y_2,\frac{\beta}{2}\Big)\bigg]}.
    \end{aligned}
    \end{equation}
    This result can also be understood as a perturbation and measurement process, albeit considerably more complex compared to the OTOC or $\mathcal{I}_3$. Here, both $\psi^1$ and $\psi^2$ induce perturbations in the past, characterized by distribution functions $h^\text{A}\Big(y_1,\frac{\beta}{2}+it_{LR}\Big)$ and $h^\text{A}\Big(y_2,0\Big)$ respectively. The perturbation from $\psi^1$ is probed by $\mathcal{U}^\dagger$, while that from $\psi^2$ is probed by both $\mathcal{U}$ and $\mathcal{U}^\dagger$. 
   
    \begin{figure}[t!]
    \centering
    \includegraphics[width=0.65\linewidth]{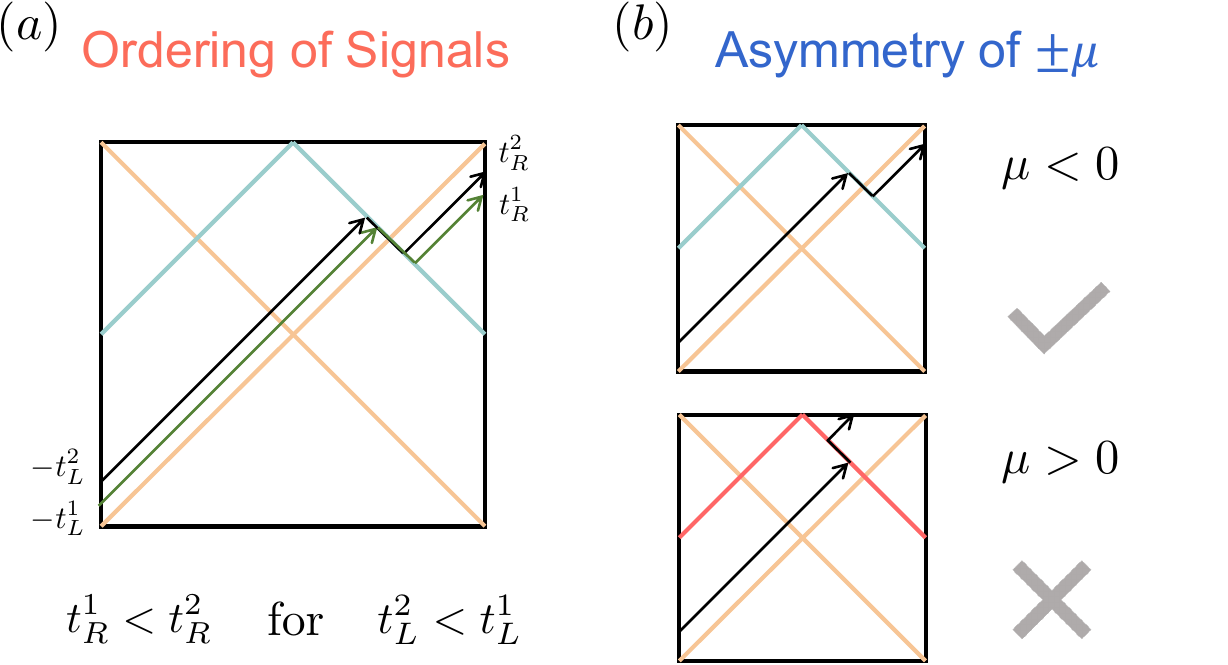}
    \caption{We illustrate signatures of semi-classical gravity, including (a) the causal ordering of two different signals and (b) the asymmetry of the signal for $\pm\mu$. }
    \label{fig: signature}
    \end{figure}

\subsection{Generalized encoding scheme}\label{subsec:generalized_encode}
    Our results can also be applied to describe more general encoding protocols in finite-size systems that may exhibit higher fidelity for wormhole teleportation. In the previous analysis, we focused on encoding information in the subsystem $L_1$, which contains Majorana fermions $\psi^1_L$ and $\psi^2_L$. Alternatively, we can opt for a more sophisticated encoding strategy by defining $L_1$ as the subsystem generated by
    \begin{equation}
    \begin{aligned}
    \Psi^1&=2^{(\mathcal{E}-1)/2}i^{\mathcal{E}(\mathcal{E}-1)/2}\psi^1\psi^3...\psi^{2\mathcal{E}-1},\\
    \Psi^2&=2^{(\mathcal{E}-1)/2}i^{\mathcal{E}(\mathcal{E}-1)/2}\psi^2\psi^4...\psi^{2\mathcal{E}},
    \end{aligned}
    \end{equation}
     with an odd integer $\mathcal{E}$. It is straightforward to check $\{\Psi^i,\Psi^j\}=\delta^{ij}$ as single Majorana fermion operators. We further choose the subsystem $R_1$ by choosing $\Psi^1_R$ and $\Psi^2_R$ that satisfies $\Psi^k_L+i\Psi^k_R|\text{EPR}\rangle$. This encoding turns out to achieve maximal fidelity in the large-$N$ limit \cite{Gao:2019nyj}. In our framework, this requires replacing all scattering vertices $\Upsilon^{\text{A},m}$ or the distribution of perturbations $h^\text{A}$ by their counterparts $\Upsilon^{\text{A},m}_\Psi$ or $h^\text{A}_\Psi$ defined for the string operator $\Psi^1$ or $\Psi^2$. By generalizing the analysis as in \eqref{eqn:composite}, we find
    \begin{equation}
    f^\text{R/A}_\Psi(x,\theta)=2^{\mathcal{E}-1}f^\text{R/A}(x,\theta)^\mathcal{E},
    \end{equation}
    if the system is invariant under an O($2p$) rotation between Majorana fermions $(\psi^1,\psi^2,...,\psi^{2\mathcal{E}})$. In the following sections, we will investigate the fidelity of teleportation for protocols with varying $\mathcal{E}$.

\section{Finite-size Effects in Teleportation}\label{sec:strong}
In this section, we analyze the results obtained in the previous section using the example of the SYK model. One of our aim is to investigate the minimum number of Majorana fermions required to observe explicit signatures of traversable wormholes, as we will clarify shortly. We also draw comparisons between strongly interacting systems exhibiting near-maximal chaos and weakly interacting systems. To study the reduced density matrix $\rho_{PR_1}$, we calculate information-theoretical quantities, including:
    \begin{enumerate}
    \item \textit{Mutual information}: The correlation between observables in $P$ and $R_1$ can be bounded by the mutual information $I(P,R_1)$:
    \begin{equation}\notag
    I(P,R_1)=S(P)+S(R_1)-S(PR_1),
    \end{equation}
    where $S(A)=-\text{tr}~[ \rho_A\ln\rho_A]$ is defined as the Von Neumann entropy of the subsystem $A$. 

    \item \textit{Negativity}: The quantum entanglement, or separability, between $P$ and $R_1$ can be probed by the negativity $\mathcal{N}(P,R_1)$:
    \begin{equation}\notag
    \mathcal{N}(P,R_1)=\frac{||\rho_{PR_1}^{T_P}||-1}{2},
    \end{equation}
    where $T_P$ represents the partial transpose relative to the subsystem $P$, and $||.||$ denotes the trace norm, which involves the summation over the absolute values of all eigenvalues.

    \end{enumerate}

    \subsection{The probe limit at infinite \texorpdfstring{$N$}{TEXT}}
    Before diving into results with finite $N$, we first explain the parameter regime for observing the semi-classical wormhole and its experimental signatures. In gravity calculations \cite{Gao:2016bin,Maldacena:2017axo}, probing a traversable wormhole without introducing significant backreactions is known as the probe limit. In quantum mechanical systems, the probe limit is defined as the parameter regime with \cite{toappear}
    \begin{equation}
    \mu\sim O(1),\ \ \ \ \ \ \ t_L\sim t_R\sim  O(\varkappa^{-1}),\ \ \ \ \ \ N\rightarrow \infty.
    \end{equation}
    Here, we only require that $\mu$ does not scale with $N$. Its value should be small, as explained in Section \ref{subsec:scramblon}. In this scenario, the coupling between the two boundaries is strong and extensive in the system size $N$, enabling teleportation within an order of $\varkappa^{-1}\ln(1/|\mu|)$ time. Focusing on this time window, we have $\lambda_0\sim O(1/N)$. Consequently, the limit of $N\rightarrow \infty$ corresponds to a Taylor expansion of $\lambda_0$ within the exponential function in Eq. \eqref{eqn:I3}, \eqref{eqn:I1_1}, and \eqref{eqn:I4}. We find 
    \begin{equation}\label{eqn:probelargeN}
    \begin{aligned}
    \mathcal{I}_3&=i f^\text{A}\left(ie^{-i\frac{\varkappa \beta}{4}}\mu N \lambda_0\Upsilon^{\text{R},1}\Big(\frac{\beta}{2}\Big),\frac{\beta}{2}+it_{LR}\right),\\
    \mathcal{I}_1&=-(\mathcal{I}_3)^2, \ \ \ \ \ \ \ \ \ \ \ \ \ \ \ \ \ \ \ \mathcal{I}_4=-\mathcal{I}_3/2. 
    \end{aligned}
    \end{equation}
    Here, the $N$ dependence cancels out in the combination $N \lambda_0=N e^{\varkappa\frac{t_1+t_2}{2}}/C$ since $C\propto N$. The result indicates that in the probe limit, the density matrix contains no more information than the linear response function $\mathcal{I}_3$. This is a generalization of discussions in Ref. \cite{Gao:2019nyj}, which is based on explicit calculations in the SYK model. This derivation can be interpreted as Majorana fermions being generalized free fields that satisfy the Wick theorem. Diagrammatically, expanding $\lambda_0$ to the leading order corresponds to introducing constraints that each source emits only a single scramblon.

    Properties of $\mathcal{I}_3$ have been elaborated in Ref. \cite{toappear}. In particular, for systems near maximal chaos $\varkappa \approx 2\pi/\beta$ \cite{Maldacena:2015waa}, the teleportation signal exhibits sharp signatures of semi-classical wormholes \cite{Jafferis:2022crx}, as illustrated in FIG. \ref{fig: signature}:
    \begin{enumerate}
    \item \textit{Causal time-order of signals}: Two signals sent into the left boundary at $t=-t^1_L$ and $t=-t^2_L$. They reach subsystem $R_1$ at times $t^1_R$ and $t^2_R$ correspondingly. Adhering to causality in the bulk, we anticipate $t^1_R<t^2_R$ given that $t^2_L<t^1_L$.

    \item \textit{Asymmetry of signals}: Teleportation necessitates negative energy generated by the coupling. In our configuration, this translates to $-i\mu\left<\psi_L\psi_R\right><0$, demanding $\mu<0$. Conversely, for $\mu>0$, the right system should not receive the teleportation signal.

    \end{enumerate}

\begin{figure}[t]
    \centering
   
    \subfloat[Mutual information]{
	\includegraphics[width=0.45\linewidth]{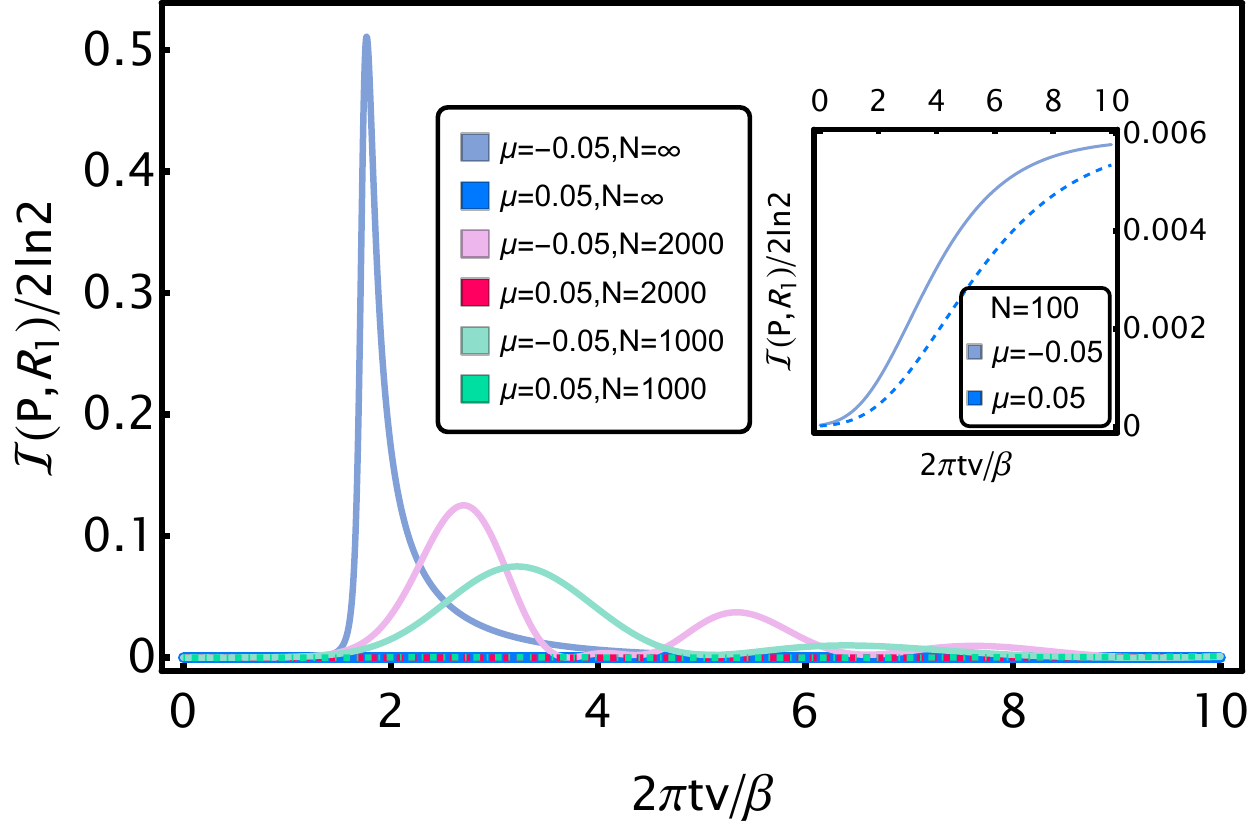}
}
\hspace{5pt}
\subfloat[Negativity]{
	\includegraphics[width=0.45\linewidth]{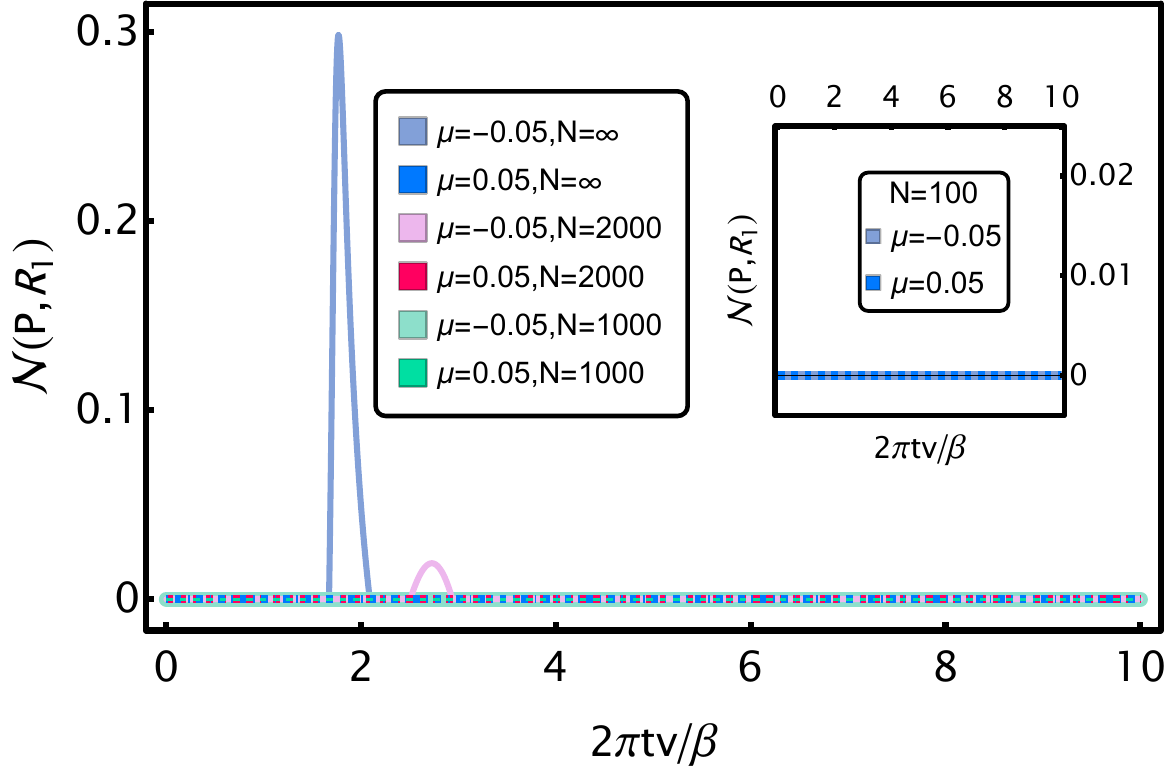}
}
    \caption{We present a comparison of (a) mutual information $I(P,R_1)$ and (b) negativity $\mathcal{N}(P,R_1)$ as a function of $t_L=t_R=t$ for different system sizes $N\in{1000,2000,\infty}$ and $\mu=\pm 0.05$ in the SYK model with $q=4$ and $v=0.95$. Insets display results for a smaller $N=100$.}
    \label{fig: num1}
    \end{figure}

\begin{figure*}[t!]
    \centering
    \includegraphics[width=0.95\linewidth]{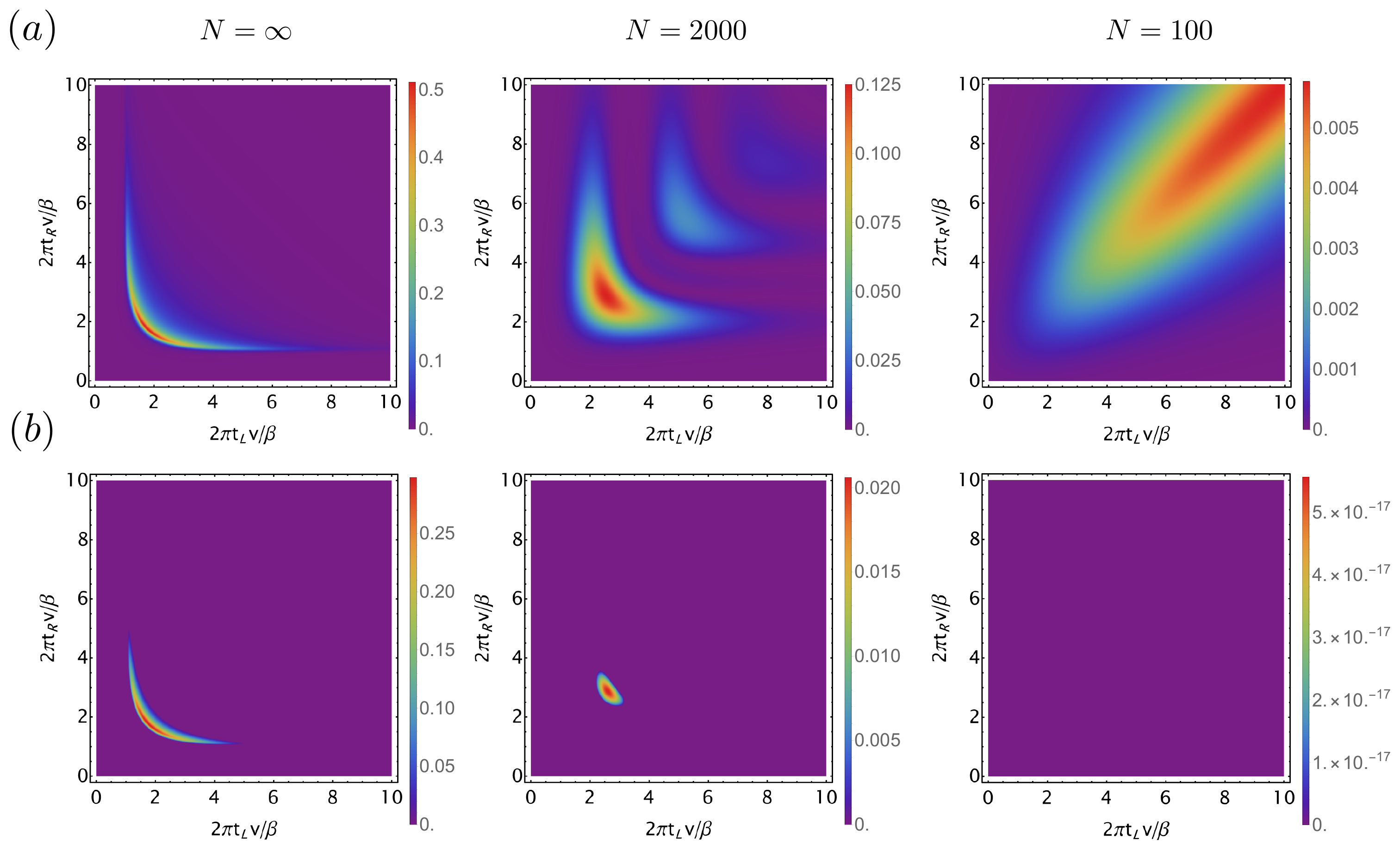}
    \caption{We provide a density plot of (a) mutual information $I(P,R_1)$ and (b) negativity $\mathcal{N}(P,R_1)$ as a function of $t_L$ and $t_R$ for different system sizes $N\in{100,2000,\infty}$ and $\mu=- 0.05$ in the SYK model with $q=4$ and $v=0.95$. }
    \label{fig: num2}
    \end{figure*}

A concrete example of holographic systems, which is also of experimental relevance \cite{Jafferis:2022crx,Luo:2017bno}, is the SYK model. This model describes Majorana fermions with random all-to-all interactions. The Hamiltonian reads \cite{Maldacena:2016hyu}
    \begin{equation}
    H=i^{q/2}\sum_{i_1\leq i_2\leq...\leq i_q }J_{i_1i_2...i_q}\psi^{i_1}\psi^{i_2}...\psi^{i_q}.
    \end{equation}
    The interaction strength $J_{i_1i_2...i_q}$ are real variables drawn from a Gaussian distribution with 
    \begin{equation}
    \overline{J_{i_1i_2...i_q}}=0,\ \ \ \ \ \ \overline{J_{i_1i_2...i_q}^2}=2^{q-1}\mathcal{J}^2(q-1)!/qN^{q-1}.
    \end{equation}
    The thermodynamics of the system is characterized by a single parameter $\beta \mathcal{J}$, and the holographic duality regime requires strong interactions/low temperatures $\beta \mathcal{J}\gg 1$. An alternative convenient parametrization is given by $\pi v/\cos (\pi v/2)=\beta \mathcal{J}$. The functions $h^\text{A}(y,\theta)$ and $f^\text{R}(x,\theta)$ have been computed in the limits of $N\rightarrow \infty$ and $\Delta=1/q\rightarrow 0$ \cite{Gu:2021xaj}
    \begin{equation}\label{eqn:largeqfh}
    \begin{aligned}
    h^\text{A}(y,\theta)&=\frac{\left(\cos \frac{\pi v}{2}\right)^{2\Delta}}{2\Gamma(2\Delta) }y^{2\Delta-1}e^{-\vartheta y},\ \ \ \ \ \ \  f^\text{R}(x,\theta)&=\frac{1}{2}\left(\frac{\cos \frac{\pi v}{2}}{\vartheta+x }\right)^{2\Delta},
    \end{aligned}
    \end{equation}
    with $\vartheta=\cos \left[v\pi\left(\frac12-\frac{\theta}{\beta}\right)\right]$. The propagator of scramblons has a prefactor $C=4\Delta^2 N\cos \frac{\pi v}{2}$ and Lyapunov exponent $\varkappa =2\pi v/\beta$. Numerical findings further confirm Eq. \eqref{eqn:largeqfh} as a reliable approximation for finite $q=4$ \cite{Gu:2021xaj}. Utilizing Eq. \eqref{eqn:largeqfh} and \eqref{eqn:probelargeN}, we plot the mutual information $I(P,R_1)$ and negativity $\mathcal{N}(P,R_1)$ in FIG. \ref{fig: num1} and FIG. \ref{fig: num2} labeled by $N=\infty$. The results clearly reveal both signatures of semiclassical wormholes. We defer the discussion of numerical results to the next subsection.

    \subsection{Looking for wormhole geometry at finite \texorpdfstring{$N$}{TEXT}}
    We are now ready to analyze how finite size systems obscure the observation of semi-classical traversable wormholes. Similar to the probe limit, we now work at the parameter regime with 
    \begin{equation}\label{eqn:strongcoupling}
    \mu\sim O(1),\ \ \ \ \ \ \ t_L\sim t_R\sim  O(\varkappa^{-1}),
    \end{equation}
    and a large but finite $N$. Therefore, we do not expand matrix elements of $\rho_{PR_1}$ in $\lambda_0$. An essential distinction arises as the system with finite $N$ manifests an interference regime in the wormhole teleportation protocol \cite{Brown:2019hmk,Nezami:2021yaq,Gao:2019nyj}. For instance, when taking the limit of $t_L\sim t_R \rightarrow \infty$, the results in Eq. \eqref{eqn:I3}, \eqref{eqn:I1_1}, and \eqref{eqn:I4} do not vanish 
    \begin{equation}\label{eqn:nnnnnn}
    \begin{aligned}
    t_L,t_R\rightarrow \infty:\ \ \ \ \ \ 
    \begin{cases}
      \mathcal{I}_1 \rightarrow e^{-i\mu NG(\frac{\beta}{2})}G\left(\frac{\beta}{2}+it_{LR}\right)^2,
    \\
    \mathcal{I}_3 \rightarrow ie^{-i\mu NG(\frac{\beta}{2})}G\left(\frac{\beta}{2}+it_{LR}\right),
    \\
     \mathcal{I}_4 \rightarrow -\frac{i}{2}G\left(\frac{\beta}{2}+it_{LR}\right).
    \end{cases}
    \end{aligned}
    \end{equation}
    In this regime, both wormhole signatures disappear for the following reasons: 
    \begin{enumerate}
        \item The matrix element reaches its maximum when $t_{LR}=0$, signifying that the correlation between $P$ and $R_1$ becomes maximal when $t_L=t_R$. 

        \item Upon taking the real part according to Eq. \eqref{eqn:real_is_important}, both mutual information and negativity exhibit symmetry for different signs of $\mu$. 
    \end{enumerate}

    To estimate the required system size $N$ for attaining the holographic regime before being influenced by the interference regime, we compare results between results with and without the expansion. The discrepancy is expected to be on the order of $\lambda_0$. Conversely, a substantial teleportation signal in the probe limit manifests when $\mu N \lambda_0\sim 1$ (refer to Eq. \eqref{eqn:probelargeN}), yielding the relationship $\lambda_0 \sim 1/(\mu N)$. Therefore, observing sharp signatures of wormholes requires $\mu N \gg 1$, which is more challenging than the naive expectation of $N\gg 1$ since the holographic picture is only justified for $\mu \ll 1$.

    Importantly, the discussion presented here also validates the application of the scramblon effective theory for probing finite-size effects. There are other diagrams that are not included in our scramblon calculation. For example, we could have
    \begin{equation}
    \begin{tikzpicture}[scale=1.15]
    \node[bvertexnormal] (R) at (-30pt,0pt) {};
    \node[sdot] (D1) at (5pt,12pt) {};
    \node[sdot] (D2) at (5pt,0pt) {};
    \node[sdot] (D3) at (5pt,-12pt) {};
    \node[bvertexsmall] (Dd) at (-5pt,6pt) {};
    \node at (2pt,15pt) {};
    \node at (2pt,0pt) {};
    \node at (2pt,-15pt) {};
    \draw[thick] (R) -- ++(135:20pt) node[left]{ };
    \draw[thick] (R) -- ++(-135:20pt) node[left]{ };
    \draw[wavy] (D1) to (Dd);
    \draw[wavy] (D2) to (Dd);
    \draw[wavy] (Dd) to (R);
    \draw[wavy] (D3) to[out=-180,in=-40] (R);
    \end{tikzpicture}\ \ \ \ \ \ \ \ \ \ \ \ \begin{tikzpicture}[scale=1.15]
    \node[bvertexnormal] (R) at (-30pt,0pt) {};
    \node[sdot] (D3) at (10pt,0pt) {};
    \node[bvertexsmall] (Dd) at (-9pt,10pt) {};
    \node at (2pt,15pt) {};
    \node at (2pt,0pt) {};
    \node at (2pt,-15pt) {};
    \draw[thick] (R) -- ++(135:20pt) node[left]{ };
    \draw[thick] (R) -- ++(-135:20pt) node[left]{ };
    \draw[wavy] (D3) to (Dd) to (R);
    \draw[wavy] (-8pt,12pt) arc (270:-90:7pt);
    \node[bvertexsmall] (Dd) at (-9pt,10pt) {};
    \end{tikzpicture}
    \end{equation}
    In comparison to diagrams in the probe limit, the contribution from the first diagram is additionally suppressed by $\mu/N$, and the contribution from the second diagram is suppressed by $1/N$. Consequently, their vanishing rates are considerably faster than those involving $1/(\mu N)$, rendering them negligible when prioritizing the summation of the most significant terms for finite-size corrections. Employing analogous reasoning, we can confidently utilize \eqref{eqn:largeqfh} in finite $N$ analysis, as their $1/N$ corrections contribute at a sub-leading level.

    With this understanding, we present the results of $I(P,R_1)$ and $\mathcal{N}(P,R_1)$ for the large-$q$ SYK model with finite $N$ in FIG. \ref{fig: num1} and FIG. \ref{fig: num2} by numerically evaluating integrals in Eq. \eqref{eqn:I3}, \eqref{eqn:I1_1}, and \eqref{eqn:I4}. Here, we set $q=4$ to align with real experiments \cite{Jafferis:2022crx} and choose $\mu=0.05$ and $v=0.95$. We have also verified that a moderate change in $\mu$ and $v$ does not alter our main conclusion. In FIG. \ref{fig: num1}, we fix $t_L=t_R=t$ and compare signals between $\mu<0$ and $\mu>0$. In mutual information, we observe a clear separation between the wormhole regime in the early-time limit and the interference regime in the long-time limit for finite but large $N\gtrsim 10^3$. Consequently, we note a distinct difference between $\pm \mu$. Additionally, the mutual information in the interference regime becomes very small, originating from the suppression of $G(\beta/2)$ at low temperatures due to a large imaginary time separation. This distinction becomes more pronounced for the negativity, which vanishes for $\mu>0$ or in the interference regime. However, when we attempt to further reduce the system size to $N=100$, the maximal mutual information in the wormhole regime quickly decreases and finally becomes comparable with the saturation value in the long-time limit. Consequently, there is no sharp distinction between an attractive coupling and a repulsive coupling. The negativity also vanishes in both cases. To examine the temporal ordering of signals, we additionally present a density plot of $I(P,R_1)$ and $\mathcal{N}(P,R_1)$ in FIG. \ref{fig: num2} for $N\in {100,2000,\infty}$. Here, a negative slope of the signal implies the preservation of causality. For $N=2000$, a negative slope can still be identified, although the peak is considerably broadened. Meanwhile, for $N=100$, we only observe a peak near $t_L=t_R$, indicative of the interference regime for teleportation. In conclusion, our results indicate $N\sim10^3$ is necessary for a justified simulation of the semi-classical traversable wormhole geometry through the wormhole teleportation protocol with a simple encoding scheme on quantum simulators.

    We comment that although dominant finite-size corrections prevent the observation of a semi-classical geometry, their contributions are still universal and closely related to gravity theory for maximally chaotic systems. The result, such as Eq. \eqref{eqn:I3}, represents an average over teleportation signals on different wormhole geometries. To see this, we introduce the distribution function $h^{\text{R}}_{\text{eff}}(y,\theta)$ as
    \begin{equation}
    e^{i\mu N[f^\text{R}(x,\frac{\beta}{2})-G(\frac{\beta}{2})]}= \int_0^\infty dy~h^{\text{R}}_{\text{eff}}(y,\theta)\exp\left(-\mu N xy\right).
    \end{equation}
    This leads to 
    \begin{equation}\label{eqn:average_geo}
    \mathcal{I}_3=i \int_0^\infty d\tilde y~ f^\text{A}\Big( \mu Ne^{-i\varkappa \beta/4}\lambda_0\tilde y,\frac{\beta}{2}+it_{LR}\Big)h^{\text{R}}_{\text{eff}}(\tilde y,\theta).
    \end{equation}
    Comparing to Eq. \eqref{eqn:probelargeN} in the probe limit, this represents an average over different wormholes with effective coupling $\mu_\text{eff}=-i\mu \tilde{y}/\Upsilon^{\text{R},1}(\beta/2)$. In the large-$N$ limit, 
    we can perform an integral using the saddle-point approximation, which fixes a single classical geometry. However, for finite-size systems, $|h^{\text{R}}_{\text{eff}}(\tilde y,\theta)|$ becomes a function with finite broadening, defining an ensemble of geometries that contributes weighted by additional phase factors. Although we expect a causal time-order of teleportation signals for fixed $\tilde{y}$, the ensemble average obscures its observation in finite-$N$ systems.

    In this section, we adopt the viewpoint that terms higher-order in $\lambda_0$ serve as perturbations in the parameter regime defined by Eq. \eqref{eqn:strongcoupling}. These terms play a dominant role in a different parameter regime with $N\rightarrow \infty$, where the coupling between the two sides is weak:
   \begin{equation}
   \mu\sim O(1/N),\ \ \ \ \ \ \ t_L\sim t_R\sim O(\varkappa^{-1}\ln N),\ \ \ \ \ \ \ N\rightarrow \infty.
   \end{equation}
   Consequently, all $N$ dependence cancels out: $\mu N\sim O(1)$ and the propagator of scramblons becomes $O(1)$. This reveals that when the coupling between the two sides is $O(1)$, one needs to wait until the information is fully scrambled before receiving the teleportation signal, similar to the discussion of the Hayden-Preskill thought experiment \cite{Hayden:2007cs}.

\subsection{Teleportation at high temperatures}
\begin{figure}[t]
    \centering
   
    \subfloat[Mutual information]{
	\includegraphics[width=0.45\linewidth]{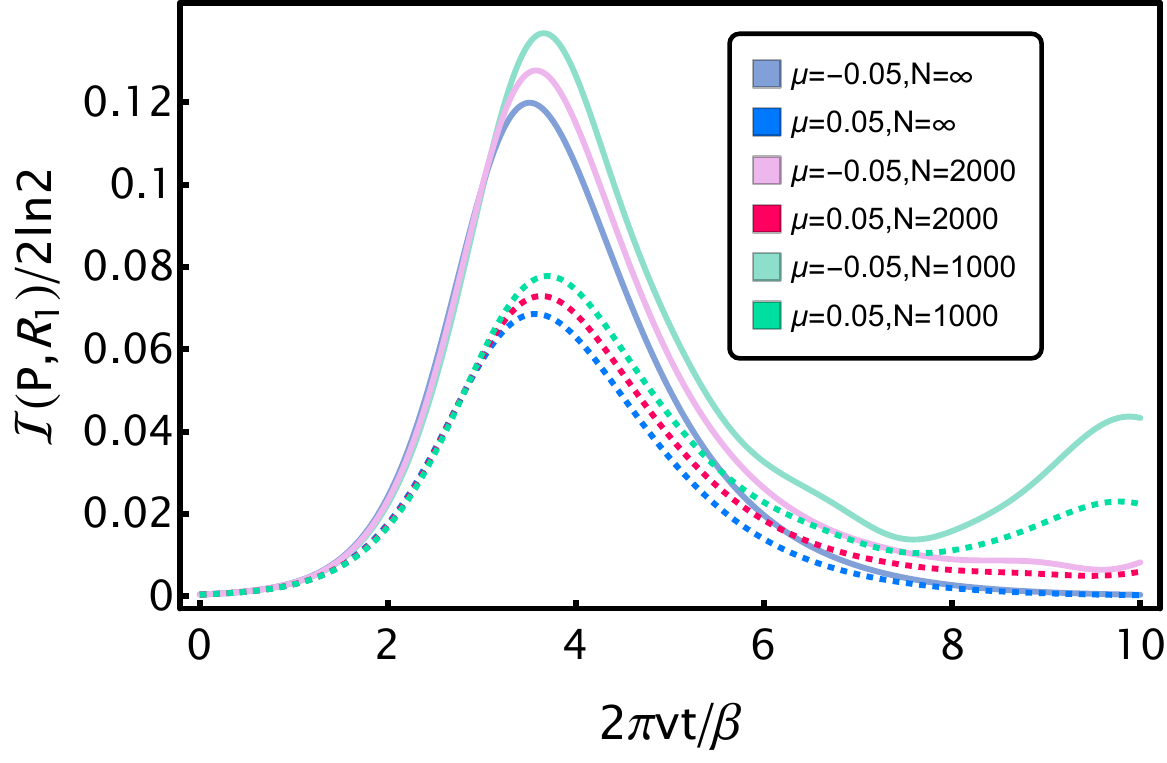}
}
\hspace{5pt}
\subfloat[Negativity]{
	\includegraphics[width=0.45\linewidth]{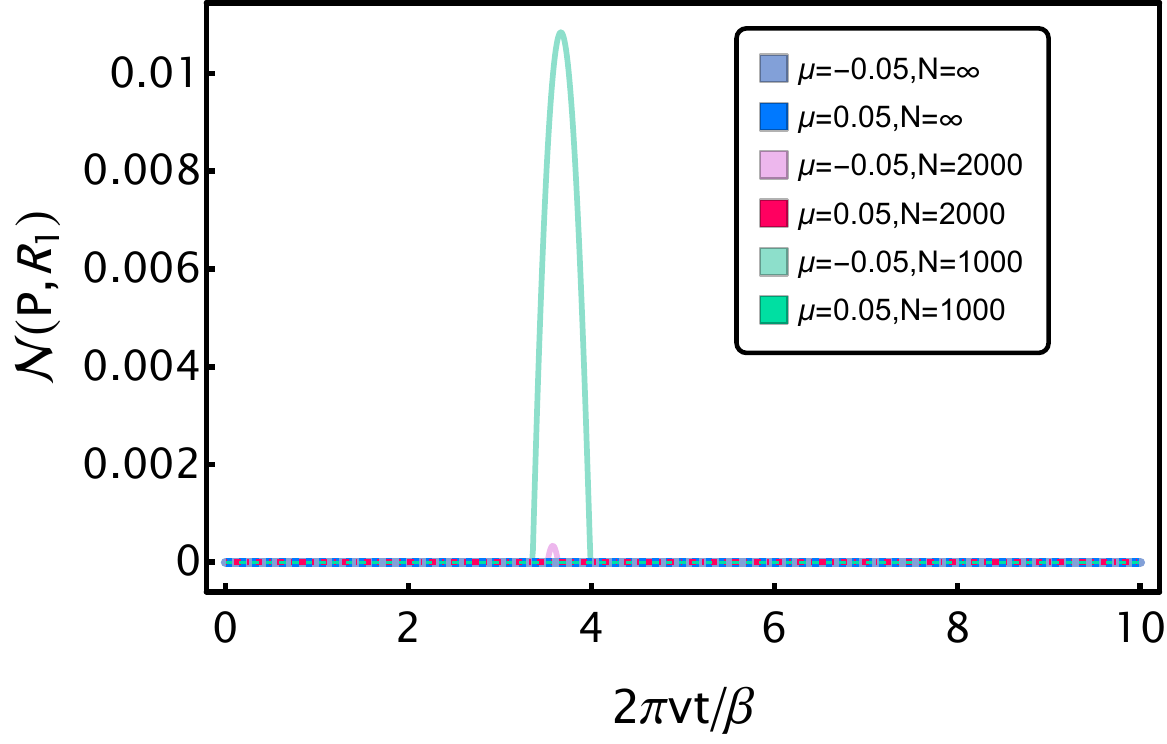}
}
    \caption{We present a comparison of (a) mutual information $I(P,R_1)$ and (b) negativity $\mathcal{N}(P,R_1)$ as a function of $t_L=t_R=t$ for different system sizes $N\in{1000,2000,\infty}$ and $\mu=\pm 0.05$ in the SYK model with $q=4$ and $v=0.1$. }
    \label{fig: num3}
    \end{figure}

 In comparison to strongly interacting systems at low temperatures, we investigate the finite-size effects of mutual information and negativity in the high-temperature limit, where the interaction is weak. Again, our focus is on the regime defined in Eq. \eqref{eqn:strongcoupling}. The numerical results for the SYK model with $q=4$ and $v=0.1$ are presented in FIG. \ref{fig: num3}. Interestingly, the results exhibit a qualitative difference compared to the low-temperature case. As we decrease $N$, there is a noticeable increase in mutual information $I(P,R_1)$. For negativity, it only becomes non-zero for finite $N$. This phenomenon can be understood by a perturbative analysis. For simplicity, we focus on the calculation of $\mathcal{I}_3$ at infinite temperature $\beta=0$ and $t_{LR}=0$. Expanding \eqref{eqn:I3} to the order of $\lambda_0^2$, we find 
   \begin{equation}\label{eqn:expansionbeta0}
   \begin{aligned}
   \mathcal{I}_3&\approx i \int dy~ h^\text{A}(y,0)e^{-i\mu N \Upsilon^{\text{R},1}(0)\lambda_0 y}\Big(1+\frac{i\mu N}{2}\Upsilon^{\text{R},2}(0)\lambda_0^2 y^2\Big)\\
&=\frac{i}{2}\left[\frac{1}{(1+p)^{2\Delta}}+\frac{i\mu N}{2}\frac{\Gamma(2\Delta+2)}{\Gamma(2\Delta)}\frac{\Upsilon^{\text{R},2}(0)\lambda_0^2}{(1+p)^{2\Delta+2}}\right].
   \end{aligned}
   \end{equation} 
   Here, we introduced $p=i\mu N \lambda_0\Upsilon^{\text{R},1}(0)$ for conciseness. The density matrix only depends on $\text{Re}[\mathcal{I}_3]$, which correpsonds to
   \begin{equation}
   \begin{aligned}
  -\frac{1}{2}\left[\text{Im}\left[\frac{1}{(1+p)^{2\Delta}}\right]+\# \mu\lambda_0^2~\text{Re}\left[\frac{1}{(1+p)^{2\Delta+2}}\right]\right].
   \end{aligned}
   \end{equation}
   As an example, we consider $\mu>0$, and $p$ lying on the positive imaginary axis. We parametrize $p=i\tan \theta$ for $\theta \in [0,\pi/2)$. The first term in the bracket is negative. It reaches the peak at $\theta_*=\pi/2(2\Delta+1)$ \cite{toappear}. We then examine the second term near the peak. It takes the value
   \begin{equation}
  \text{Re}\left[\frac{1}{(1+i \tan \theta_*)^{2\Delta+2}}\right]=-\sin \theta_*(\cos\theta_*)^{2(\Delta+1)}<0.
   \end{equation}
   This analysis is independent of $\Delta$. Therefore, the finite-size correction due to the second term takes the same sign as the first term, leading to an overall increase in the total signal. A comparable analysis can be conducted for systems approaching maximal chaos. However, in this case, the second term in \eqref{eqn:expansionbeta0} incorporates an extra factor of $e^{-i\varkappa \beta/2}\approx -1$. Consequently, the teleportation signal diminishes rapidly.

\subsection{Encoding in Majorana strings }
\begin{figure}[t]
    \centering
   
    \subfloat[Mutual information]{
	\includegraphics[width=0.45\linewidth]{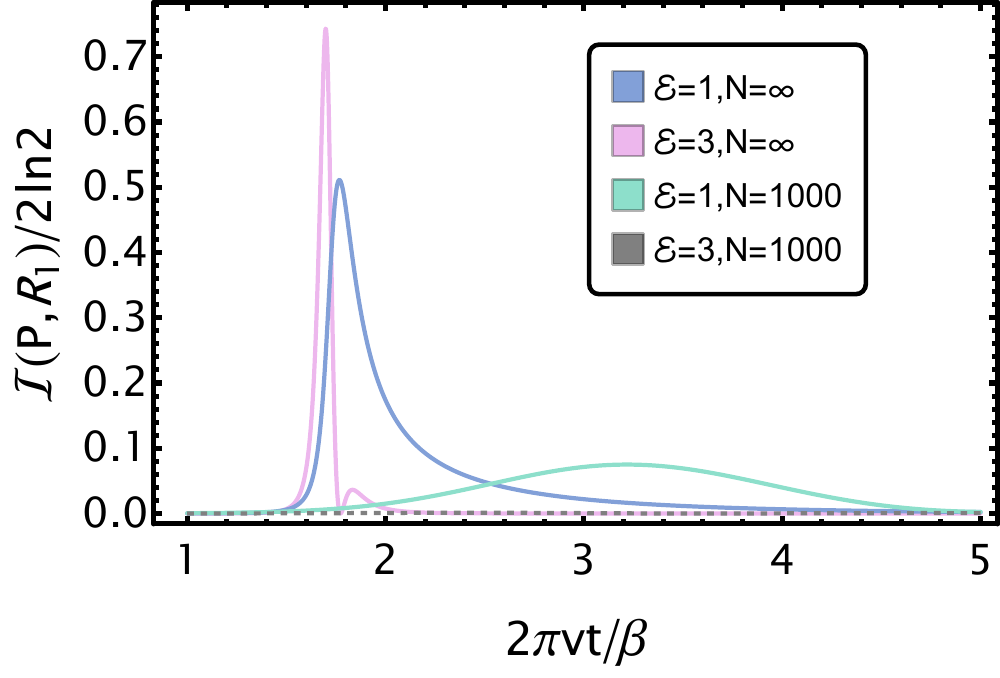}
}
\hspace{5pt}
\subfloat[Negativity]{
	\includegraphics[width=0.45\linewidth]{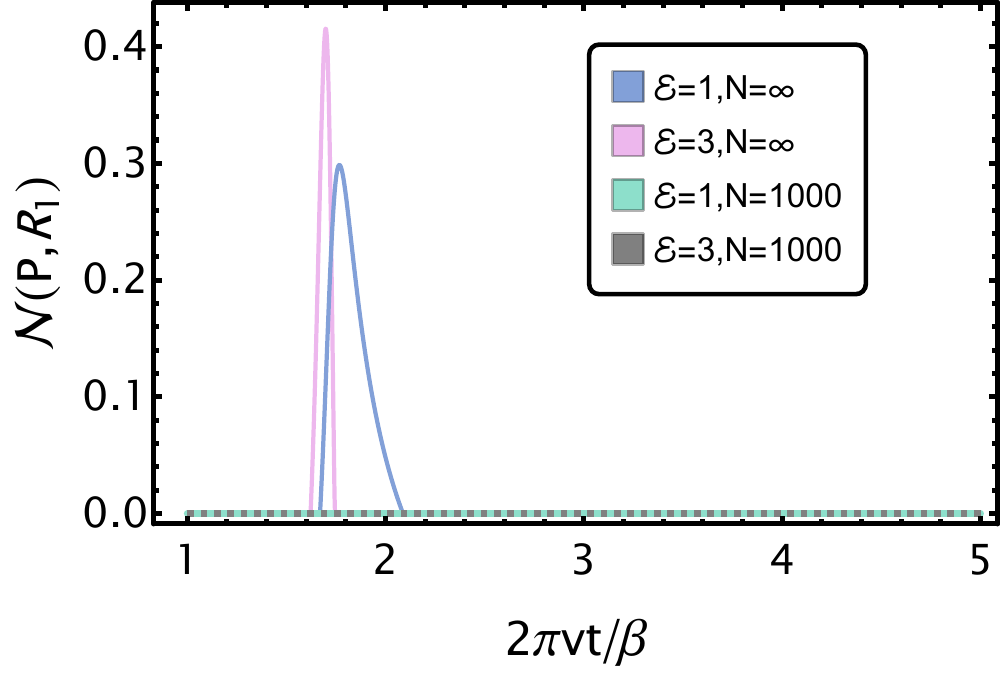}
}
    \caption{We present a comparison of (a) mutual information $I(P,R_1)$ and (b) negativity $\mathcal{N}(P,R_1)$ as a function of $t_L=t_R=t$ for different system sizes $N\in{1000,\infty}$ and $\mathcal{E}=1,3$ in the SYK model with $q=4$, $\mu=-0.05$ and $v=0.95$. }
    \label{fig: num4}
    \end{figure}

In Ref. \cite{Gao:2019nyj}, the authors propose that the fidelity of wormhole teleportation can approach its maximum (in the probe limit) when considering the generalized encoding scheme in section \ref{subsec:generalized_encode}. It is then natural to ask whether a similar trick can be employed to reduce finite-size effects. Unfortunately, we find the answer to be negative. As discussed in section \ref{subsec:generalized_encode}, the only difference is to replace the advanced scattering vertices by $\Upsilon^{\text{A},m}_\Psi$. Therefore, Eq. \eqref{eqn:average_geo} becomes
   \begin{equation}
    \mathcal{I}_3=i 2^{\mathcal{E}-1}\int_0^\infty d\tilde y~ f^\text{A}\Big( \mu Ne^{-i\varkappa \beta/4}\lambda_0\tilde y,\frac{\beta}{2}+it_{LR}\Big)^\mathcal{E}h^{\text{R}}_{\text{eff}}(\tilde y,\theta).
    \end{equation}
    When $\mathcal{E}$ increases, two effects come into play: 1. the maximum of $f^\text{A}$ can become larger, as discussed in \cite{Gao:2019nyj}; 2. the peak becomes narrower since $f^\text{A}( x,\theta )^\mathcal{E}\sim 1/x^{2\Delta \mathcal{E}}$ for large $x$. Therefore, although effect 1 dominates as $N\rightarrow\infty$, the teleportation signal decreases more rapidly compared to $\mathcal{E}=1$ when the system size $N$ decreases due to larger boardening, leading to a more severe finite-size effect. Numerical demonstration is provided in FIG. 7.

    \section{Summary}
    
    In this work, we establish a general framework for describing wormhole teleportation in chaotic systems with all-to-all interactions. The calculation takes into account finite-size effects, providing a natural crossover between the wormhole regime and the interference regime. Using the concrete example of the SYK model, we observe that the teleportation signal, characterized by mutual information and negativity, decreases rapidly in the low-temperature limit as $N$ decreases. We further estimate that the observation of a single semi-classical wormhole geometry requires $N\sim 10^3$. Otherwise, sharp signatures generally get smoothed out when averaged over different geometries. In comparison, the signal gets amplified for large but finite $N$ in the high-temperature limit, as justified in a perturbative analysis.

    Several comments are as follows: Firstly, in addition to the original SYK model, various generalizations exhibiting maximal chaotic behavior have been proposed, such as those in \cite{Fu:2016vas,Gu:2016oyy,Banerjee:2016ncu,Chen:2017dav,Jian:2017unn,Peng:2017kro,Gu:2019jub}. One may wonder whether some of these models allow for simulations with smaller system size $N$. As an example, we may expect the finite-size effect to be suppressed when $\Delta$ is small, due to the presence of the factor $\frac{\Gamma(2\Delta+2)}{\Gamma(2\Delta)}$ in Eq. \eqref{eqn:expansionbeta0}. In the original SYK model, achieving a small $\Delta$ requires a larger $q$, which is unfavorable as it increases the complexity of simulating the Hamiltonian dynamics. However, a small $\Delta$ can be achieved in the bipartite SYK model \cite{Fremling:2021wwy}, which only contains four-body couplings. This provides a possibility for reducing the required $N$. Secondly, our analysis also offers an opportunity for experimental demonstration of the intriguing relation between operator size growth and wormhole teleportation. In \cite{Liu:2023lyu}, the authors developed an experimental approach for probing the distribution functions $h^\text{R/A}$ by measuring the operator size distribution in random spin models. Our results then indicate a relation between the operator size distribution and the fidelity of quantum teleportation. Finally, it is known that there are transitions for information scrambling when the system is embedded in an environment \cite{Zhang:2022knu}. It would be interesting to investigate the consequences on wormhole teleportation. 
    Finally, previous studies \cite{Brown:2019hmk,Nezami:2021yaq} have identified size winding as the mechanism for wormhole teleportation, which can be measured experimentally \cite{Jafferis:2022crx}. Our calculations suggest that size winding with a holographic origin also receives significant corrections at finite $N$. A detailed analysis is presented in a companion manuscript \cite{Tian-GangZhou:2024rcd}, employing the scramblon effective theory in the large-$q$ SYK model. The results confirm our expectation: with finite size corrections, the winding size distribution near the scrambling time contains a complex phase factor (see Eq. (18) in \cite{Tian-GangZhou:2024rcd}), which is generally non-linear in the operator size. On the other hand, in the long-time limit, although the system exhibits size winding, the teleportation is non-holographic, as observed in \eqref{eqn:nnnnnn}. Therefore, observing size winding with a holographic origin is also challenging.

\vspace{5pt}

\textit{Acknowledgment.}
We thank Ping Gao, Yingfei Gu, Cheng Peng, Xiao-Liang Qi, Huajia Wang, Jinzhao Wang, Zhenbin Yang, and Tian-Gang Zhou for helpful discussions.
This project is supported by the NSFC under grant 12374477. 

\vspace{5pt}
\textit{Note added.} 
When completing this work, we became aware of an independent investigation of teleportation fidelity by computing the quantum channel capacity in 2D Jackiw-Teitelboim gravity with backreactions \cite{newpaper}.

\bibliography{draft.bbl}

\end{document}